\documentclass[twocolumn,aps,prb,floats]{revtex4}
\usepackage{graphics,graphicx,epsfig}
\usepackage{amssymb}
\usepackage{epsf,epstopdf,wrapfig}

\begin{document}

\title{Adaptive Filtering Enhances Information Transmission in Visual Cortex}

\author{Tatyana O. Sharpee$^{1,2}$, Hiroki Sugihara$^2$, Andrei
  V. Kurgansky$^2$, Sergei P. Rebrik$^2$, 
Michael P. Stryker$^{1,2}$, and Kenneth D. Miller$^{1-3}$}

\affiliation{$^1$Sloan-Swartz Center for Theoretical Neurobiology and
  $^2$Department of Physiology, \\ University of California, San
  Francisco, 513 Parnassus Avenue, San Francisco, CA 94143-0444. \\ 
  $^3$Center for Theoretical Neuroscience, Center for Neurobiology and
  Behavior, \\Columbia University Medical School, N.Y.S.P.I. Kolb
  Research Annex, 1051 Riverside Drive, Unit 87, New York, NY
  10032-2695. }

\date{February 23, 2006}

\begin{abstract}
  Sensory neuroscience seeks to understand how the brain encodes
  natural environments. However, neural coding has largely been
  studied using simplified stimuli.  In order to assess whether the
  brain's coding strategy depend on the stimulus ensemble, we apply a
  new information-theoretic method that allows unbiased calculation of
  neural filters (receptive fields) from responses to natural scenes
  or other complex signals with strong multipoint correlations. In the
  cat primary visual cortex we compare responses to natural inputs
  with those to noise inputs matched for luminance and contrast. We
  find that neural filters adaptively change with the input ensemble
  so as to increase the information carried by the neural response
  about the filtered stimulus.  Adaptation affects the spatial
  frequency composition of the filter, enhancing sensitivity to
  under-represented frequencies in agreement with optimal encoding
  arguments. Adaptation occurs over 40~s to many minutes, longer
  than most previously reported forms of adaptation.
\end{abstract}

\maketitle

The neural circuits in the brain that underlie our behavior are well
suited for processing of real-world -- or natural -- stimuli. These
neural circuits, especially at the higher stages of neural processing,
may be largely or completely unresponsive to many artificial stimulus
sets used to analyze the early stages of sensory processing and, more
generally, for systems analysis. Thus, natural stimuli may be
necessary to study higher-level neurons. Characterizing neural
responses to natural stimuli at early or intermediate stages of neural
processing, such as the primary visual cortex, is a necessary step for
systematic studies of higher-level neurons. Neural responses are also
known to be highly nonlinear\cite{Theunissen00,Sceniak02,Nolt04} and
adaptive\cite{Maffei73,Shapley79,Shapley_Enroth_84,Ohzawa85,Saul89,Smirnakis97,Brenner00-adapt,Dragoi00,Fairhall01,Chander01,Baccus02,Kohn04,Solomon04,Victor87,Brown01,Movshon79,Albrecht84},
making them difficult to predict across different stimulus
sets\cite{David04}.  Therefore, even early in visual processing,
characterizations based on simplified stimuli may not be adequate to
understand responses to the natural environment.

For these reasons there has been a great deal of interest in studying
neural responses to complex, natural stimuli (for example, see refs
\cite{Theunissen00,David04,Baddeley97,Theunissen01,Ringach02,Smyth03,Felsen05}).
However, the relationship between coding of natural and laboratory
stimuli remains elusive due to the difficulty of characterizing
neurons -- assessing their receptive fields -- from responses to
natural stimuli, as we now describe.

A simple and commonly-used model of neural responses is the {\em
  linear-nonlinear}  model\cite{deBoer,Rieke_book}. In this model,
the response of the neuron depends on linear filtering of the stimulus
luminance values {\bf S} by a receptive field {\bf L} defined over
some region of space and time. Mathematically, the filter output at
time $t$ is a sum over the spatial positions ($x$,$y$) and temporal
delays $t'$ to which the neuron's response is sensitive:
$\sum_{x,y,t'}${\bf L}$(x,y,t-t')${\bf S}$(x,y,t')$, which we
abbreviate as {\bf L}*{\bf S}. The output of this filter is then
passed through a nonlinear function $f$ to yield the neuron's response
$r$: $r(t) = f(${\bf L}*{\bf S}$)$. The nonlinearity incorporates the
fact that the firing rate cannot be negative and other aspects of
neural response such as threshold, saturation, and sensitivity or
insensitivity to changes in stimulus polarity. We will use the terms
neural filter or receptive field throughout this paper to mean the
linear part {\bf L} of the linear-nonlinear model.

Traditionally, neural receptive fields have been estimated as the
spike-triggered average stimulus (STA; with appropriate correction
for autocorrelation of the
inputs)\cite{Theunissen00,Theunissen01,Ringach02,Smyth03,deBoer,Rieke_book}
or by related methods\cite{Brenner00-adapt,Felsen05,Rust05}.  These
methods give unbiased results for linear systems for any stimulus
ensemble or for nonlinear systems if the ensemble is Gaussian random
noise. However, they produce systematic deviations from the true
filter of nonlinear ``linear-nonlinear'' neurons probed with natural stimuli (or other
non-Gaussian stimuli), even in situations where the only nonlinearity
is due to a conversion of the output of a linear receptive field to
firing rate\cite{Ringach02,Sharpee04}. This happens because natural
stimuli, unlike Gaussian stimuli which may be completely described by
pairwise correlations, have strong higher-order as well as pairwise
correlations\cite{Ruderman,Field94,Simoncelli01}. The higher-order
correlations may be viewed as what distinguishes natural from random
Gaussian stimuli. The bias in the filter estimate calculated using the
Gaussian or linear assumption increases with the strength of the
nonlinearity and with the strength of stimulus correlations beyond
second order\cite{Ringach02,Sharpee04}, not vanishing even with
infinite data.

Recently an information-theoretic method has been developed that
correctly estimates receptive fields of nonlinear model neurons (with
extensions to multiple linear filters) for arbitrary stimulus
ensembles regardless of the strength of multi-point correlations, even
in cases where the STA is zero\cite{Sharpee04}. According to this
method, one searches for the spatiotemporal filter {\bf L} whose
output, {\bf L*S}, carries the most mutual information with the
experimentally measured neuronal response $r(t)$.  In practice, this
is done via a gradient ascent procedure, searching in the space of all
possible spatiotemporal receptive fields or filters to find the most
informative one (referred to as ``the most informative dimension'', or
MID). We can then calculate the nonlinearity associated with the MID
from the data as the probability of a spike given the filter output;
there is no need to make any assumption about the shape of the
nonlinearity.

Similarly to other ``spike-triggered'' methods, the MID method compares
two probability distributions of outputs for a given filter: the
distribution of outputs that occur before (or trigger) a spike, and
the distribution of outputs over the entire stimulus ensemble
regardless of neural response. If a filter represents a stimulus
feature that affects neural responses, then certain values of its
output will be more probable before a spike, and so the two
distributions should differ from one another. The various methods all
seek filters that maximize the difference between the two
distributions, but differ in the measure of this difference.  For the
STA, the measure is the change in the mean of the two distributions;
for the spike-triggered covariance
method\cite{Brenner00-adapt,Felsen05,Rust05}, it is the change in the
variance; and for the MID, it is an information-theoretic measure (the
Kullback-Leibler distance) that corresponds to the mutual information
between the filter output and the spikes. The information-theoretic
measure is more general than the mean or variance, because it is
sensitive to correlations of all orders, which in part explains the
success of the MID method in estimating neural filters from responses
to natural stimuli. Here we apply this method for the first time to
neural data, focusing on the single-filter model, to address the
question of whether and how V1 receptive fields adapt to natural
stimuli.

\section{Receptive fields from noise vs. natural scenes}

\begin{figure*}
\includegraphics[width=7in]{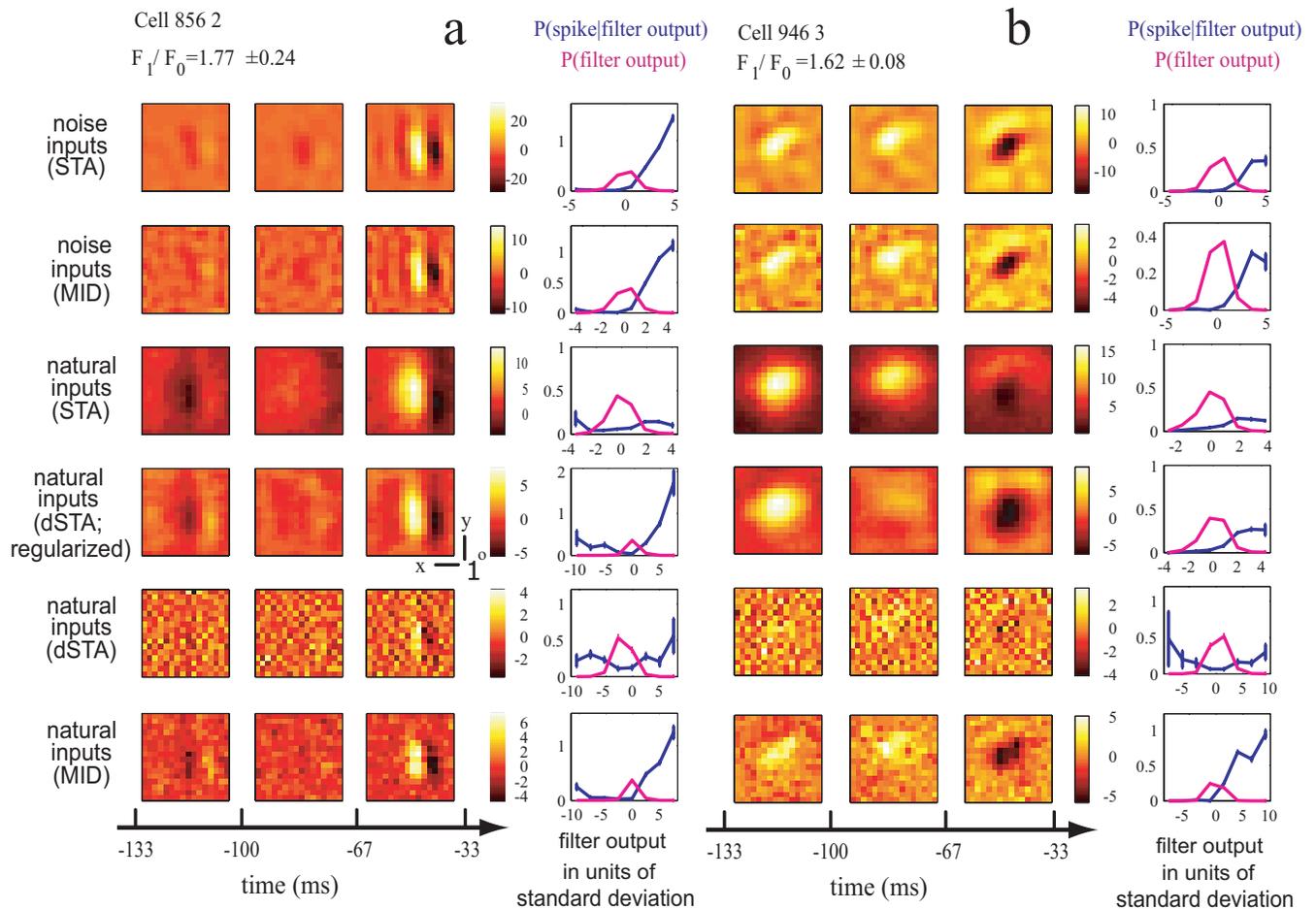}
\caption{{\bf Filters and nonlinearities for two simple cells.}  Top
  to bottom: STA and MID for noise ensemble; STA, dSTA, dSTA with
  regularization, and MID for natural ensemble. Spatiotemporal
  receptive fields have three time frames covering the indicated
  interval (-133 to -33 ms). In the right-most column for each filter
  we plot the probability distribution of filter outputs in the
  stimulus ensemble (magenta) and the spike probability given the
  filter output (blue; values of the $y$ axis refer to these
  probabilities). The color scale shows the filter in units of its
  average noise level (see Supplementary Methods). $x$-$y$ scale bars:
  1$^o$. Error bars show standard errors of the mean in all figures.
}
\end{figure*}

We studied 40 simple cells (as characterized by responses to optimal
moving gratings\cite{Skottun91}) in anesthetized cat V1 (complex cells can also be
characterized by the MID method\cite{Sharpee04} and will be considered
in a future publication). We probed these neurons with natural and
white noise inputs. These inputs differ in two important respects.
First, they have very different pairwise correlations, which are
described by the power spectra. The power spectrum of a white noise
ensemble does not depend on either spatial or temporal frequency
within a certain range, while the power spectrum of natural inputs
depends on spatial frequency $k$ as $\sim 1/k^2$ under a wide variety
of conditions\cite{Ruderman,Simoncelli01,Field87,Dong95} (spatiotemporal
statistics have similar structure\cite{Dong95}). Second, natural scenes have
strong statistical correlations beyond second order that cannot be
described by the power spectrum, as evident for example in the much
greater incidence of oriented edges in natural scenes than in Gaussian
noise with the same power spectrum\cite{Ruderman,Simoncelli01}.

To estimate spatiotemporal receptive fields or neural filters from
responses to noise and natural stimuli, we applied both the linear
systems and information-theoretic methods. The resulting estimated
filters and STAs for two example cells are shown in Fig. 1. With
respect to responses to the noise ensemble, we found the filter for
each cell either as the traditional STA or as the MID\cite{Sharpee04}.
As expected for white noise stimuli, the two estimates do not differ
significantly from each other for the illustrated cells or for most
cells ($p>0.05$ for 31 out of 40 cells; $t$-test, see Supplementary
Methods); the remaining differences can be attributed to the residual
spatial correlations in the white noise ensemble (cf. Fig. 3b). This
agreement illustrates the basic validity of the MID method under
circumstances where the STA offers an independent unbiased estimate.

For responses to the natural stimulus ensemble, we calculated the STA
and corrected it for second-order correlations present in the natural
ensemble to obtain a decorrelated STA (dSTA). This would describe the
neuron's filter if the neuron were linear. Because this procedure of
correcting for stimulus correlations tends to amplify noise, we also
calculated the dSTA using regularization to prevent such amplification
-- such decorrelation with regularization has been used in most
previous work estimating neural filters from responses to natural
signals\cite{Theunissen00,David04,Ringach02,Smyth03,Felsen05}.
Finally, we estimated the filter from natural inputs as the MID. As
can be seen in Fig. 1, the MID produces an estimate of the filter for
natural scenes that is much closer to the white noise filter than
either the dSTA or the regularized dSTA.  Across cells, the dSTA shows
a greater difference from the white noise filter than does the natural
ensemble MID, as judged by smaller correlation coefficients with
either the noise ensemble STA or noise ensemble MID (40/40 cells,
$p<10^{-6}$). This demonstrates that some of the differences between
the neural filters obtained from natural and noise stimulation in the
linear model are due to biases in the estimation of the natural filter
that can be removed once the linear-nonlinear model is considered and
the MID is computed. In Fig. 1, we also plot the nonlinear functions
that show spike probability as a function of filter output. They are
similar in shape for the MIDs of the two ensembles, and this behavior
seems to be typical across cells.

We used the MIDs to estimate both the noise and natural filters in
what follows. We studied all simple cells with a non-zero filter to
both natural and noise inputs.
        
Despite the similarity of the filters obtained under the two
conditions, cf. Fig. 1, a jackknife analysis of the errors in
estimating the neural filters shows that the differences between the
filters derived from noise and natural signals are statistically
significant ($p<0.01$) for all cells. To investigate the source of
these differences and to make connections with classic studies on
neural responses to moving periodic patterns (gratings) of certain
orientations and spatial frequencies, we compute the spatiotemporal
Fourier transform of the filter in the two spatial dimensions and
time. The position of the maximum of the Fourier transform at the
grating temporal frequency is our prediction for the optimal grating
orientation and spatial frequency for a particular neuron. We did not
detect any systematic shifts in optimal orientation and only a small
shift in optimal spatial frequency as assayed from noise filters,
natural signals filters and grating stimuli, in agreement with
previous findings using the regularized dSTA\cite{Smyth03,Felsen05},
see Supplementary Discussion.

\begin{figure*}
\includegraphics[width=5in]{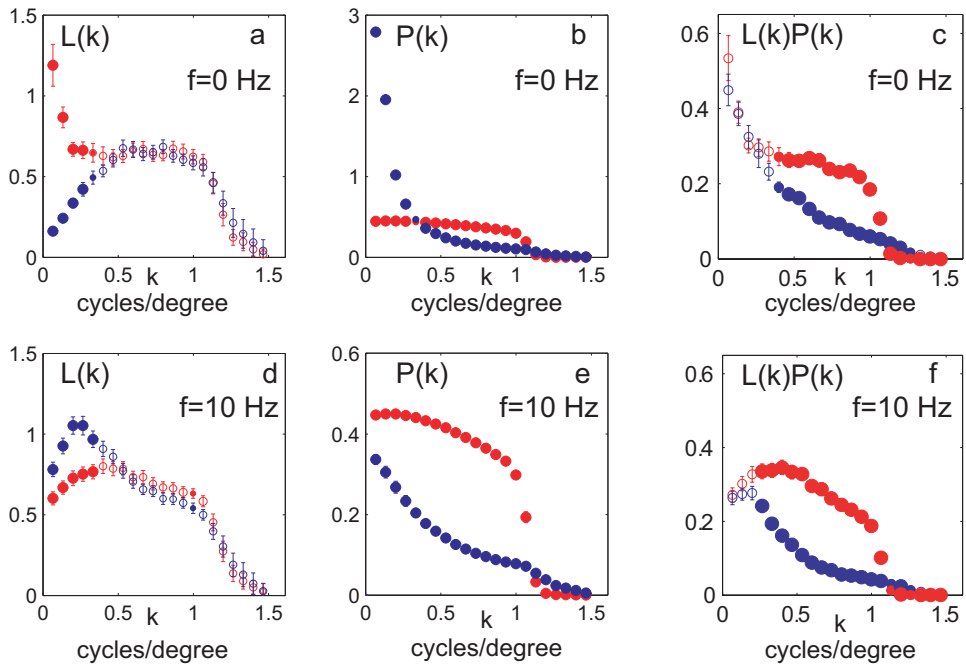}
\caption{{\bf Neural filters compensate for changes in the input power spectrum.}
  Average amplitude spectra of neural filters (a,d) and input
  ensembles (b,e) corresponding to natural (blue circles) and white
  noise (red circles) stimulation for temporal frequencies of 0 and 10
  Hz. The spectra were taken along the optimal orientation for each
  cell by interpolating the discrete 2D Fourier transform. We use
  filled circles at frequencies where mean sensitivity was
  significantly different between the two ensembles (small circles for
  $p<0.05$ and large for $p<0.01$), and open symbols otherwise. (c, f)
  Plots of the product of the average neural filter and input ensemble
  amplitude spectra.  }
\end{figure*}

The most marked differences between the neural filters derived from
natural vs. noise stimulation are seen by considering the entire shape
of the spatial frequency tuning curves (Fig. 2 and Supplementary Figs
1 and 2) and not just the location of the single best spatial
frequency. For each cell and temporal frequency, we calculated the
spatial frequency profile along the cell's preferred stimulus
orientation using interpolation of the filter's two-dimensional
discrete Fourier transform. Note that our temporal resolution allowed
analysis only at two temporal frequencies (0 Hz and 10 Hz, in each of
two opposite directions of motion). Results at 10Hz did not depend on
direction of motion so both directions were combined in Fig. 2, which
shows the average tuning of the cells in our dataset. For low spatial
frequencies sensitivity decreased (increased) to common (rare) inputs,
while at middle and high spatial frequencies the sensitivity did not
change. For example, at zero temporal frequency, low spatial
frequencies are more common in the natural than in the white noise
stimulus ensemble (Fig. 2b). Correspondingly, neurons became less
sensitive to those frequencies during stimulation with natural inputs
than during stimulation with noise inputs (Fig. 2a). In the case of
non-zero temporal frequencies the trend is reversed, because the noise
stimulus ensemble has more power at nearly all spatial frequencies
than the natural stimulus ensemble (Fig. 2d, e). These changes in
filter can be observed in the majority of cells, and are not simply
due to adaptation in a small subset of cells. This is shown in
Supplementary Fig. 1, which illustrates the spatial frequency
sensitivities of the two example cells whose receptive fields are
shown in Fig. 1, and Supplementary Fig. 2, which shows scatter-plots
of spatial frequency sensitivity of noise vs. natural filters across
all cells.

\section{Optimal filtering in a nonlinear system.}

In retrospect, such shifts in spatial frequency sensitivity may be
expected for neural coding to be optimal for both of two input
ensembles (whit noise and natural stimuli) that have such vastly
different power spectra as white noise and natural
stimuli\cite{Ruderman,Simoncelli01,Field87} (see Fig.  2b, e). In
general it is difficult to map optimal coding strategy from one
ensemble to another; however, it could be done if both of the stimulus
ensembles were Gaussian so that they were entirely characterized by
their power spectra. Suppose a neuron uses filter $L_A$ and
nonlinearity $f_A$ to optimally encode Gaussian stimulus ensemble $A$
with spatiotemporal amplitude spectrum $P_A(k,\omega)$.  What would
then be an optimal strategy to encode Gaussian ensemble $B$ with
amplitude spectrum $P_B(k,\omega)$? One solution is to leave the
nonlinearity unchanged and to compensate for differences in the input
power spectra by changing neural filter properties so that:

\begin{equation}
L_A(k,\omega)\cdot P_A(k,\omega)=L_B(k,\omega)\cdot P_B(k,\omega)
\end{equation}

This will leave unchanged all statistics of neuronal response, and so
in particular will leave invariant any statistical measures of
optimality.  Alternative strategies involving a change in nonlinearity
cannot be optimal unless there are multiple optima, because if
ensemble $A$ has a unique optimum, then the above strategy will give
the unique optimum for ensemble $B$.  (Note that, in response to an
overall change in contrast, the nonlinearity can be
rescaled\cite{Brenner00-adapt,Fairhall01}, but this is equivalent to a
rescaling of the filter according to Eq. 1 with no change in
nonlinearity.)

These conclusions about the receptive field and nonlinearity apply
only to Gaussian stimuli. The higher-order correlations present in
natural scenes may both lead to deviations from Eq. (1) in neural
filters and cause changes in the shape of the nonlinearity. But in
practice, the changes in the shape of the nonlinearity are small, and
changes in neural filters that do take place act to compensate for
changes in the input power spectrum as predicted from Eq. (1) (Fig.
2c,f). These changes in frequency sensitivity occur primarily at low
spatial frequencies. No changes are observed at mid-to-high spatial
frequencies, resulting in significant deviations from Eq. (1) in the
middle range of frequencies.  We can only speculate that other factors
may limit the range of frequencies over which adaptation can occur.

\section{Adaptation Increases Information Transmission}

The above optimal coding argument provides at least a qualitative
explanation of observed receptive field changes. Most theories of
optimal coding define optimality in information-theoretic terms. To
test directly whether the information maximization argument applies to
our data, we calculated the average mutual information between the
filter output and the neural response; the response at a given time is
simply taken as the presence or absence of a single
spike\cite{Sharpee04}.

\begin{figure}
\includegraphics[width=3.4in]{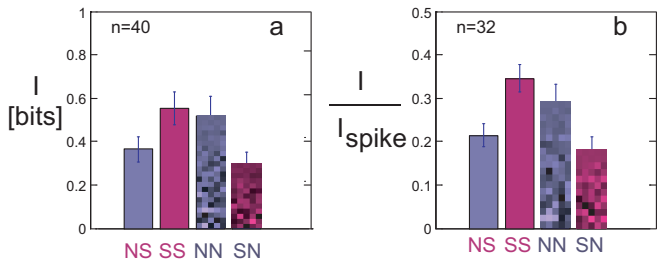}
\caption{{\bf Receptive field adaptation increases information transmission.}
  Bars show the mutual information between spikes and outputs of
  either noise (blue, N) or natural scenes (red, S) filter applied to
  natural scenes ensemble (solid) or noise ensemble (pixelated). NS,
  white noise filter applied to natural scenes ensemble; SS, natural
  scenes filter applied to natural scenes ensemble; NN, white noise
  filter applied to noise ensemble; SN, natural scenes filter applied
  to noise ensemble. The information values are given in bits (a) or
  in units of the total information carried by the arrival of a single
  spike $I_{spike}$\cite{B00a} (b).  }
\end{figure}

The changes in receptive fields act to increase the information after
changes in stimulus ensemble, and this information would be
substantially reduced if receptive fields did not change with the
ensemble. That is, the natural filter carries more information about
responses to the natural ensemble than to the noise ensemble
($p<10^{-4}$, paired Wilcoxon two-tailed test), whereas the noise
filter carries more information about responses to the noise ensemble
than to the natural ensemble ($p=0.03$). The average information
values across the population are shown in Fig. 3, and scatter-plots on
a cell-by-cell basis are provided in Supplementary Fig. 4. Each filter
produces roughly equal information about responses to its own
ensemble: the difference in information values achieved by applying
the noise filter to the noise ensemble versus applying the natural
filter to the natural ensemble is not significant ($p=0.18$, paired
Wilcoxon test). Each filter produces substantially less information
about responses to the other ensemble ($p<10^{-4}$ for natural or
noise ensemble filtered with natural versus noise filter; paired
Wilcoxon tests), and there is no significant difference between the
swapped combinations (natural filter applied to noise ensemble or visa
versa, $p=0.06$, paired Wilcoxon test). We note that the changes in
information are not due to overfitting or other computational
artifacts, because information was calculated from responses to
ensemble segments that were not used in calculating the filters, and
the effects were not seen in data from a model linear-nonlinear cell
with unchanging filter that was analyzed similarly, see Supplementary
Information.

In addition to considering information $I$ in bits (Fig. 3a), we also
measured information for each cell in units of $I_{\rm
  spike}$\cite{B00a}, the information in the neuron's response (as
defined above) about the full stimulus (Fig. 3b). $I/I_{\rm spike}$
measures the fraction of the total possible information that is
captured by the single most informative filter ($I_{\rm spike}$ is a
separate measurement that was available only for a subset of cells,
making the data set smaller). As can be seen, the MID captures roughly
35\% of the possible information for simple cells.  Each filter
provides a greater fraction of the overall information when applied to
its own ensemble than the other ($p\leq 10^{-4}$ for natural filter
applied to natural vs.  noise ensemble and for either ensemble
filtered with natural vs. noise filter; $p=0.05$ for noise filter
applied to natural vs. noise ensemble; paired Wilcoxon test).

\section{Dynamics of receptive field adaptation}

Even though the best linear-nonlinear model systematically changes
with the stimulus ensemble, this does not establish that the neuron
has changed its encoding strategy. The true encoding strategy may be
complicated and nonlinear, so that even if it is static, the best
linear-nonlinear estimate of it may change with the ensemble, much as
the best linear approximation to a curve changes with position on the
curve.

\begin{figure*}
\includegraphics[width=5in]{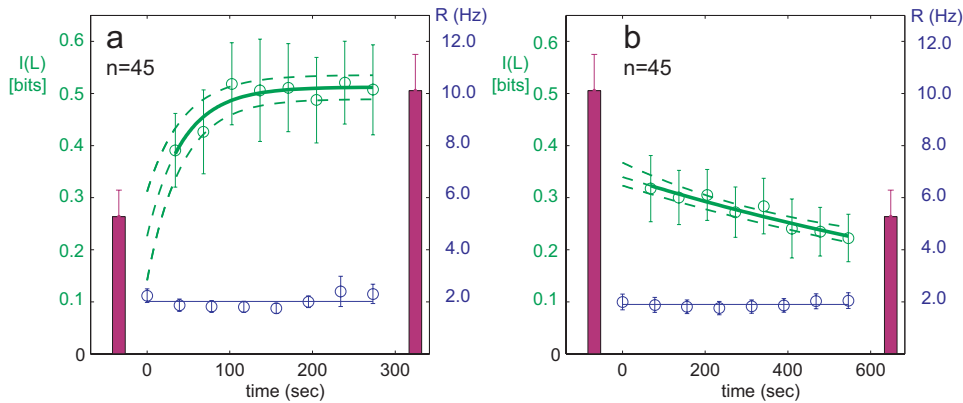}
\caption{{\bf Adaptation dynamics.}
a,b,  The neural filter derived from the last half of natural stimulation
  is applied to the first half of natural (a) or to the noise (b)
  ensemble. Symbols show information (green, left $y$ axis) and firing
  rate (blue, right $y$ axis) averaged across cells, versus time. The solid
  line is an exponential fit; dashed lines show one standard deviation
  based on the Jacobian of the fit [$p=0.01$ in a and $p=0.003$ in b
  using an $F$-test with null hypothesis of no time dependence]. The taller
  (shorter) red bars show information for the natural filter applied
  to natural (noise) inputs (as in Fig.~3, but $n=45$). The firing
  rates demonstrate that recordings are stable.  }
\end{figure*}

The most direct method to distinguish between an adaptive strategy and
a complex but static coding strategy would be to estimate the filter
as a function of time and see it change.  This method yields very poor
time resolution, because $\sim 5$~min of data are needed to
estimate the filter, so adaptation that occurs on a faster timescale
cannot be seen.  Nonetheless we tried this method and saw appropriate,
if weak, adaptation to noise stimuli even on this long time scale (see
Supplementary Fig. 5).  To achieve finer time resolution, we studied
adaptation by measuring changes with time in the information carried
by the output of a single, static filter; this information can be
estimated from $\sim 30$~s of data. We used the following
reasoning.  If the coding is static, then the mutual information
between this filter's output and the neuron's responses to a given
ensemble should not systematically change in time.  However, if the
neuron's receptive field adapts to the stimulus ensemble, then this
information may systematically change in time.  In particular, we take
the static filter to be that characterizing a neuron when it is well
adapted to a given ensemble -- say the natural ensemble.  When the
neuron is newly exposed to a natural ensemble, the information carried
by this filter should increase with increasing time of exposure to the
natural ensemble, as the neuron adapts so that the filter that it
actually uses to encode incoming stimuli into spikes becomes closer
and closer to this static, fully adapted filter.  Similarly, when the
neuron is newly exposed to a noise ensemble, the information carried
by this filter should decrease with increasing time of exposure to the
noise ensemble, as the neuron's own filter adapts to the noise and
becomes less and less like the fully adapted natural scenes filter.

We derived filters from the last half of the 10-min presentations of
each stimulus ensemble, when the neuron would be best adapted to the
given ensemble if adaptation occurs. We then applied these static
filters to both noise and natural stimuli, and measured information
between spikes and filtered stimuli in successive 34-s periods
during the first half of stimulus presentation (if the filter was
derived from the second half of this stimulus) or in successive 68-sec
periods during all of the presentation of the opposite ensemble. Most
cells did not show significant adaptation when considered
individually, presumably due to the variability in measuring
information over such brief time periods.  However, averaging over the
entire population of simple cells revealed clear adaptation over time,
consistent with an adaptive coding strategy (Fig. 4). The information
progressively increased with time when natural inputs were filtered
with the neural filter derived from the natural stimulus ensemble
(Fig. 4a; see also Supplementary Discussion and Supplementary Fig.
6), while the information decreased with time when that same filter
was applied to noise inputs (Fig. 4b).

Fits of a single exponential to the average data demonstrate that
there is a statistically significant monotonic change with time, with
time constants $\tau =42\pm 9$~s for adaptation to the natural
ensemble and $\tau=22\pm 2 $~min for adaptation to the noise ensemble.
These time constants are consistent with the fact that we could not
detect adaptation to the natural ensemble with the 5-min time scale
of direct filter measurements, but we could detect adaptation to the
noise ensemble (Supplementary Fig. 5).  Note, however, that the time
constants are based on the assumption of exponential decay, and do not
exclude the possibility of multiple time scales, including scales
faster than we were able to measure, or of alternative functional
forms of decay.

We could not detect a significant trend with time in the information
carried by the noise filter about either ensemble (see Supplementary
Fig. 7). This is perhaps not surprising given that the average
decrease in information for the noise filter applied to the noise
versus natural ensembles was not significant ($p=0.14$, unpaired
$t$-test), and that the slow time course of adaptation to the noise
ensemble suggests that the filter we tested was not fully adapted to
it (see also legend of Supplementary Fig. 7).  Nonetheless, the
presence of significant monotonic changes in the expected directions
for the natural scenes filter applied to each ensemble demonstrates
that the neuron's coding strategy is adapting over time with exposure
to a given ensemble.

\section{Discussion}

Adaptation is ubiquitous throughout the nervous system, and it occurs
in many forms. In vision, adaptation to luminance mean and variance
(contrast) has been observed in the
retina\cite{Shapley79,Shapley_Enroth_84,Smirnakis97,Chander01,Baccus02,Victor87,Brown01},
lateral geniculate nucleus\cite{Solomon04} and primary visual
cortex\cite{Maffei73,Ohzawa85,Saul89,Albrecht84}, and related changes
are observed in perception\cite{Blakemore69}. In the framework of our
model, adaptation may affect the neural gain (the nonlinear
input-output function), or the spatiotemporal filter itself.
Adaptation of the gain to the mean and variance of the stimulus
ensemble (and perhaps to higher-order statistics\cite{Kvale}) serves
to fit a neuron's dynamic range to the dynamic range of the
stimulus\cite{Shapley79,Shapley_Enroth_84,Ohzawa85,Smirnakis97,Brenner00-adapt,Fairhall01,Chander01,Baccus02,Solomon04}.
In addition, adaptation of the filter to the mean and variance of the
stimulus\cite{Sceniak02,Nolt04,Shapley79,Shapley_Enroth_84,Baccus02,Victor87}
has been observed, and it has been argued that such adaptation along
with adaptation to the stimulus covariance can serve to maximize the
information per spike in the neuron's response\cite{Wainwright,Atick}.
In general, filter adaptations are nearly instantaneous ($<0.1$~s),
while changes in gain can be more gradual (time constants up to 10~s,
and perhaps longer for some components of adaptation to mean
luminance)\cite{Shapley79,Shapley_Enroth_84,Smirnakis97,Chander01,Solomon04,Victor87}.
Here we find an adaptive change in neural filters in response to
stimulus statistics beyond the mean and variance, and one that occurs
over much longer time scales than previously found even for contrast
gain changes. This suggests that the observed adaptation represents a
new mechanism for optimal coding.

Adaptation to the power spectrum could be considered a generalized
form of contrast adaptation, in which different frequency channels
providing input to cortical cells differentially adapt their gains so
that channels with more stimulus power show greater adaptation.
Indeed, variation of gain adaptation across different retinal pathways
has been observed\cite{Smirnakis97,Chander01,Solomon04,Brown01}.
However, these observations, and a recently reported pattern-specific
component of retinal adaptation\cite{Hosoya05}, involved adaptation on
significantly faster time scales than observed here. Also, in the
lateral geniculate nucleus, adaptive changes between white noise and
natural stimulation were not observed in the temporal domain, at least
for a majority of cells\cite{Dan96}. This suggests that the adaptive
changes reported here are of cortical origin. A pattern-specific
component of cortical adaptation has been observed: for example, one that
differentially affects responses according to the difference of the
stimulus orientation, direction, or spatial frequency from that of the
adapting stimulus\cite{Saul89,Dragoi00,Kohn04,Movshon79,Albrecht84}.
At least in one case, this adaptation has been observed to have time
constants on the order of a minute or longer\cite{Dragoi00}. It is
possible that the present observations may share some underlying
mechanisms with such pattern-specific adaptation.

Many recent studies have used versions of the linear model or related
models to estimate receptive fields from responses to natural
stimuli\cite{Theunissen00,David04,Theunissen01,Ringach02,Smyth03,Felsen05}.
Some have reported that the estimates calculated from responses to
natural stimuli differ from those calculated from responses to
noise\cite{Theunissen00,David04,Theunissen01}, whereas
others\cite{Smyth03,Felsen05} found no change in the major parameters of
neural filters, such as optimal stimulus orientation and spatial
frequency. It is not clear from these observations to what degree
reported differences in neural filters are genuinely stimulus-induced
or are due to biases in the estimation induced by the non-Gaussian
statistics of natural stimuli together with the nonlinearity of the
input-output function. The fact that the receptive field obtained for
a given ensemble from the linear model best predicted responses to
other examples of its own
ensemble\cite{Theunissen00,David04,Theunissen01} suggests at least
partially genuine differences, which is also supported by our results
on spatial frequency adaptation. However, the fact that we found
larger differences between filters obtained in the linear
approximation (dSTA for natural stimulus ensemble and STA for white
noise ensemble) than between filters obtained in the linear-nonlinear
model (MID for natural stimulus ensembles and STA or MID for white
noise ensemble) suggests that biases also exist, and the new
information maximization procedure used here removes these biases for
real neurons, just as was demonstrated in numerical
simulations\cite{Sharpee04}.

We have found that V1 neurons adapt their filters to stimulus
statistics beyond the mean and variance. This filter adaptation occurs
over 40~s to many minutes, suggesting it is not a consequence of
previously described mechanisms of luminance or contrast adaptation.
The adaptation serves to preserve information transmission and to
reduce relative responses to stimulus components that are relatively
more abundant in the stimulus ensemble, as predicted by optimal
encoding arguments. It remains to be determined whether the neurons
are adapting to changes in power spectra, in higher-order statistics,
or both. The gradual nature of adaptive changes and their
correspondence to optimization principles suggests that it might be
possible to predict the direction and degree of adaptation to stimulus
sets with statistics intermediate between those of white noise and
natural stimuli. Thus, there is hope for creating a unified picture of
neural responses across various input ensembles.

\section{Methods}

All experimental recordings were conducted under a protocol approved
by the University of California, San Francisco on Animal Research with
procedures previously described\cite{Emondi04}. Spike trains were
recorded using tetrode electrodes from the primary visual cortex of
anesthetized adult cats and manually sorted off-line. Visual stimulus
ensembles of white noise and natural scenes were each 546~s long.
After manually estimating the size and position of the receptive
field, neurons were probed with full-field moving periodic patterns
(gratings). Cells were selected as simple if, under stimulation by a
moving sinusoidal grating with optimal parameters, the ratio of their
response modulation ($F_1$, that is amplitude of the Fourier transform
of the response at the temporal frequency of the grating) to the mean
response ($F_0$) was larger than one\cite{Skottun91}. The rest of the
protocol typically consisted of an interlaced sequence consisting of
three different noise input ensembles of identical statistical
properties, and three different natural input ensembles. The interval
between presentations varied in duration as necessary to provide
adequate animal care. All natural input ensembles were recorded in a
wooded environment with a hand-held digital video camera in similar
conditions on the same day, see Supplementary Movie. The noise
ensembles were white overall, but the spatial frequency spectrum was
divided into eight circular bands, and each particular frame was
limited to one band at random; this white noise design was intended to
increase the number of elicited spikes. The mean luminance and
contrast of the noise ensembles were adjusted to match those of the
natural ensembles. Both noise and natural inputs were shown at 128x128
pixel resolution, with angular resolution of approximately 0.12$^o$
per pixel. To calculate receptive fields, input ensembles were
down-sampled to 32x32 pixels. The receptive field center was
determined from the maxima in the STAs for noise and natural ensembles
and was set to the same position for analysis of both noise and
natural inputs. A patch of 16x16 pixels was selected around the center
(angular resolution of 0.48$^o$ per pixel) to make analysis
computationally feasible and to minimize effects due to undersampling
(we strove to have the number of spikes greater than the
dimensionality of the receptive fields\cite{Sharpee04}). In all cases
subsequent analysis of receptive fields verified that the selected
patch fully contained the receptive field. These receptive fields were
used in all quantitative analyses, Figs 2-4. Examples in Fig. 1 were
computed at and are shown at twice the angular resolution to
illustrate the finer structure of the receptive field, as well as
differences in performance of the various methods.

\section{Acknowledgments}

We acknowledge suggestions from W. Bialek on the design of experiments
and subsequent data analysis. We thank M.  Caywood, B. St. Amant , and
K. MacLeod for help with experiments. We thank P. Sabes, M. Kvale and
S. Palmer for many helpful suggestions on statistical aspects of data
analysis. Computing resources were provided by the National Science
Foundation under the following NSF programs: Partnerships for Advanced
Computational Infrastructure at the San Diego Supercomputer Center
through NSF cooperative agreement ACI-9619020, Distributed Terascale
Facility (DTF) and Terascale Extensions: enhancements to the
Extensible Terascale Facility.  This research was supported through
grant R01-EY13595 to K.M. from the National Eye Institute and by a
grant from the Swartz Foundation and a career development award
K25MH068904-02 from the National Institutes of Mental Health to T.S.

Correspondence and Requests for materials should be addressed to:
sharpee@phy.ucsf.edu


\section{Supplementary Iinformation.}
\subsection{Supplementary Discussion.}

{\bf Optimal filtering in a nonlinear system.} The simple argument
leading to Eq. (1) may appear reminiscent of the redundancy reduction
principle\cite{Wainwright,Atick,Barlow61,Barlow01,Atick90}. However we
do not assume that the response is linear, impose a particular
constraint, or specify the optimality measure. We simply assume that
the optimality measure, whatever it may be, is preserved under a
change in ensemble.  Due to the generality of this argument, we cannot
make predictions for the optimal shape of frequency tuning, only for
the relative changes in tuning upon a change in the input power
spectra. For a linear system, the redundancy reduction arguments
predict\cite{Wainwright,Atick,Atick90} that neural filters should
completely remove second-order correlations present in the input
ensembles, i.e. the product $L(k)P(k)$ should be constant across
frequencies for sufficiently small frequencies for any ensemble.
Although this argument may reasonably describe subcortical visual
processing\cite{Atick,Dan96,Atick90}, it does not appear to describe
visual cortex either in response to natural stimuli or to noise (Fig.
2c,f), where $L(k)P(k)$ depends on $k$.  Therefore nonlinearities of
simple cells and/or alternative optimization principles appear
essential in describing optimal filter properties in the primary
visual cortex.

\begin{figure*}
\includegraphics[width=5in]{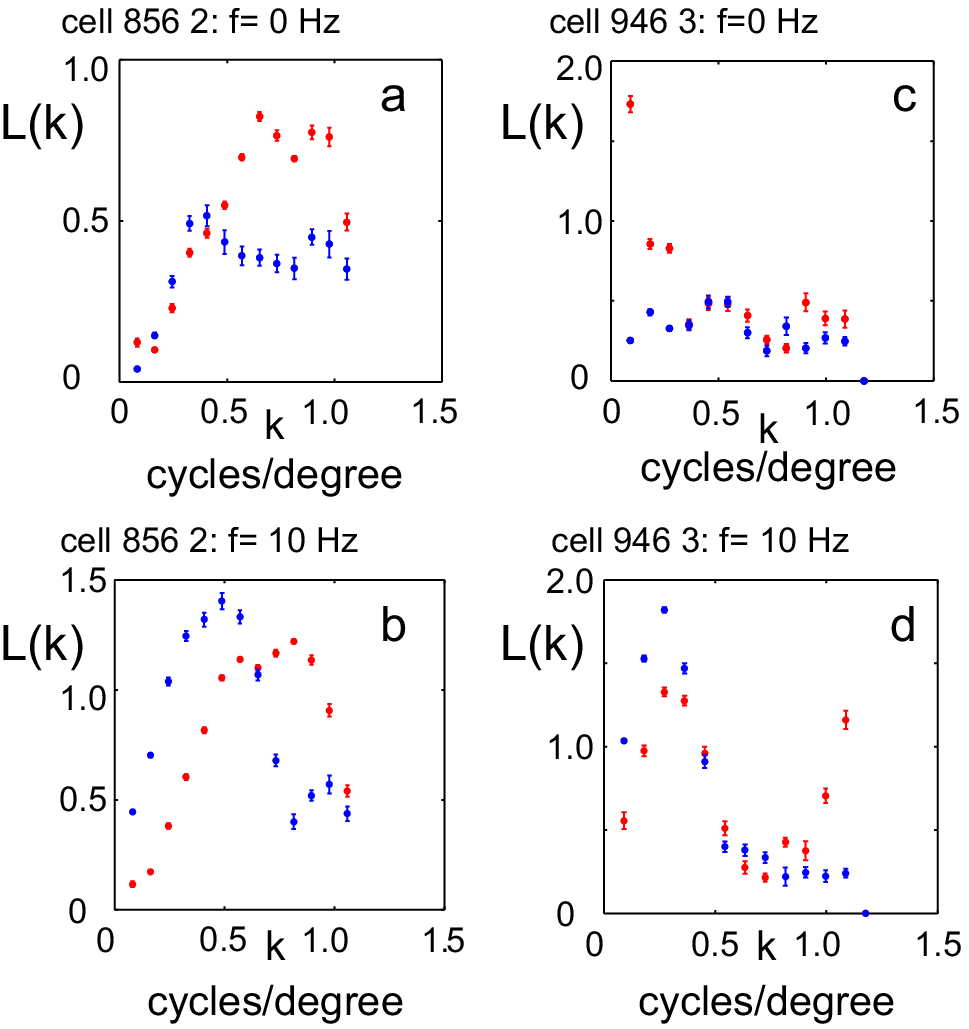}
\parbox{7.0in}{
\begin{flushleft}
SUPPLEMENTARY FIG. 1: This figure shows the spatial frequency profiles of receptive fields
  from the two example cells of Fig.~1. Spatial frequency
  sensitivity at zero temporal frequency (a,c) and at 10 Hz (b,d).
  Red indicates filter derived from responses to noise ensemble, blue
  indicates filter derived from responses to natural ensemble. The
  second of the two example cells is typical in all respects. The
  first of the two cells is atypical in that it did not change its
  sensitivity at low spatial frequency between natural and noise
  stimulation at 0 Hz, but exhibited an appropriate change in its
  tuning at 10 Hz, see Supplementary Fig.~2.
\end{flushleft}}
\end{figure*}

In the Discussion of the main text, we point out that the adaptation
observed here may share some underlying mechanisms with previous
observations of cortical pattern-specific adaptation. Indeed, it has
been proposed\cite{Wainwright,Barlow90,Barlow89} that such
pattern-specific adaptation arises from anti-Hebbian or decorrelating
mechanisms that would more generally lead to adaptation to the
stimulus power spectrum like that observed here. These models of
adaptation\cite{Wainwright,Barlow90,Barlow89} are closely related to
the redundancy reduction arguments just discussed and, more generally,
to principles of optimal
encoding\cite{B00a,Atick,Barlow61,Barlow01,Atick90} that have been
proposed to govern the design and operation of the nervous system.
Despite the specific disagreements just discussed, our results support
these general ideas in two respects. First, we have found that
adaptation acts to reduce relative responsiveness to patterns that
have relatively greater stimulus power, as these theories predict.
Second, we have found that neural filters adapt to changes in stimulus
ensemble in a manner that increases the information transmitted,
relative to the information that would be transmitted if filters did
not adapt (as seen by the decreased information when filter and
ensemble are swapped, Fig. 3).

The optimality argument (1) for a nonlinear system analyzing Gaussian
inputs predicts that the nonlinearity does not change its functional
form. This is supported in our data by the fact that the average
information values are roughly equal under natural and noise
stimulation. The information $I$ can be rewritten in terms of the
nonlinear function $f(x)$ of the filter output $x$ and the probability
$P(x)$ that the filter output has value $x$: $I = \int dx
P(x)f(x){\rm log}_2f(x)$.  One way to preserve this sum is to use the
strategy of our optimality argument: to leave $P(x)$ unchanged, which
for a Gaussian ensemble is accomplished by changing the filter
according to Eq. 1, and to leave the nonlinearity $f(x)$ unchanged. The
extent to which this strategy is followed by our two example cells can
be seen in Fig.~1 and Supplementary Fig.~3: the pink curves
illustrate $P(x)$, while the blue curves illustrate $f(x)$, which in
Fig.~1 is also scaled by the firing rate. As can be seen by
comparing the curves for the noise MID to that for the natural MID,
the curves are at least roughly preserved.

{\bf Sensitivity of simple cells to multiple stimulus dimensions.}  We
find that even simple cells in primary visual cortex are sensitive to
more than one stimulus dimension, in agreement with other recent
work\cite{Felsen05,Rust05}.  A single filter corresponds to a single
stimulus dimension; the filter output tells the strength of the
stimulus along that dimension.  The ratio $I/I_{\rm spike}$ (Fig. 3b)
tells the proportion of the information encoded about the stimulus in
the neuron's spikes that can be accounted for by the output of the
most informative filter\cite{Sharpee04}.  For simple cells, the
dominant filter accounts for only about 35\% of the overall
information. Thus, other stimulus dimensions must significantly
influence the neuron's firing\cite{Brenner00-adapt,Felsen05,Rust05}.
Presumably all of these relevant dimensions also shift with changes in
stimulus ensemble, of which we analyzed here only the dominant one. It
is also possible that adaptive changes in the structure of each of the
relevant dimensions will change their relative importance for
eliciting a spike. In particular, the dominant filter for one input
ensemble might become secondary in encoding the other input ensemble.
The fact that we did not see qualitative changes in the structure of
the dominant filter between natural and noise stimulation suggests
that such shifts in the relative role of dimensions are not common.
Future studies will extend the adaptation analysis to include other
relevant dimensions beyond the dominant filter.

{\bf Optimal spatial frequency and orientation under natural and noise
  stimulation.}  Filters derived from noise and natural stimuli had
similar optimal orientation and spatial frequency. The optimal values
were obtained as the position of the maximum of the 2D Fourier
transform in space at the temporal frequency of the grating (2Hz).  We
found a small but statistically significant shift in the optimal
spatial frequency, with filters derived from noise inputs having a
21\% ($\pm$~3\% s.d.)  higher value of the optimal spatial frequency
than filters derived from natural inputs ($p<10^{-4}$). This shift in
optimal spatial frequency was small enough that neither the noise
ensemble estimate nor the natural ensemble estimate was significantly
different from direct measurements of the preferred spatial
frequencies of these cells with gratings. We note that the
measurements with gratings were done separately, before exposure to
the noise or natural ensembles, and do not represent tests of grating
spatial frequency sensitivity in the states of adaptation to white
noise or natural stimuli.  We note also that our conclusions about
optimal coding depend on the sensitivity throughout the entire range
of spatial frequencies and not on the position of the maximum of the
spatial frequency tuning curve (the ``optimal" spatial frequency) for a
particular cell.

In agreement with previous findings\cite{Smyth03,Felsen05}, we did not
see statistically significant changes in optimal stimulus orientation
between grating, natural ensemble, or noise ensemble estimates.
Natural stimuli have anisotropic power spectra with increased power at
horizontal and vertical orientations\cite{Coppola98}, and therefore
one might have expected some shifts in optimal stimulus orientation
away from horizontal or vertical for the natural filter relative to
the noise filter. Adaptation to orientation is strongest when the
difference between the preferred orientation of the neuron and the
adapting orientation is between 20-60 degrees, and acts to shift the
preferred orientation away from the adapting stimulus\cite{Dragoi00}. Thus, shifts
due to over-representation of vertical and horizontal orientations
would both tend to occur on neurons preferring oblique orientations,
and would be in opposite directions. We speculate that the two effects
tend to cancel.

{\bf Dynamics of Adaptation to Natural Stimuli.} Here we argue against
certain artifactual explanations of Fig.~4a. It could be argued that
the increase of information with time seen in Fig.~4a may occur
because of correlations between the stimuli used for the information
calculation (the ``test set'') and those used in calculating the
filters themselves (the ``training set''). Natural movies tend to have
correlations that diminish in time as a power law rather than an
exponential\cite{Ruderman,Simoncelli01,Field87,Dong95}, and in that
sense are long-lasting. The training set was the last half of the
movies, so it might be argued that, as time progresses from the
beginning of the movies, the correlation of the test set with the
training set would increase and this might explain the increase in
information. One argument against this explanation is that information
saturates after the first quarter of stimulus presentations, whereas
the correlation with the training set would continue to increase
throughout the first half. We tested this explanation more directly by
using an alternative training set.  We calculated the filters from the
middle half of the movies (136 to 410 sec) and then calculated
information on the first quarter and the last quarter.  Now the first
quarter and the last quarter are equally distant in time from the
training set, and so if this explanation were correct we would expect
them to be mirror images of each other: information would go up during
the first quarter and go down by an equal amount during the last
quarter. On the contrary, and in support of the adaptation argument,
we see the same rise in information during the first quarter as
before, even though the first quarter is now much closer in time to
the training set, and we see no fall in information during the last
quarter, cf. Supplementary Fig.~6a. An exponential fit gave a time
constant of $55\pm 9$~s, which agrees with the time constant of $42\pm
9$~s derived from information during the first half of the data, cf.
Fig.~4.  Also, against the more general argument that the rise or fall
in information in Fig.~4 might be due to some non-stationarity in the
stimulus movies, we show that relevant stimulus components, such as
the mean and the standard deviation of the outputs of the neural
filters applied to these movies, are stable, cf. Supplementary Fig.~6
(b-e).

\subsection{Supplementary Methods.}
{\bf Dataset Selection.}  The present dataset is obtained from 4
animals and included 133 single units which were clustered using a
manual spike sorter. For 85 of the 133 neurons, a reliable non-zero
filter was obtained from natural inputs, as judged by visual
inspection. We found that this subjective criterion correlated well
with an objective criterion of having a significantly positive
information value for the filter applied to its own ensemble (after
finite-size corrections\cite{entropy} are applied). The information
was positive for all 85 cells, and exceeded its standard deviation in
81/85 cells.  We used the latter criterion to select the dataset of 71
cells with reliable filter estimates to both noise and natural
stimuli, of which 40 were classified as simple based on their
responses to moving sinusoidal gratings of optimal orientation and
spatial frequency.  Specifically, simple cells were those with ratio
of $F_1/F_0>1$, where $F_1$ is the response modulation (Fourier
component at the frequency of the stimulus grating) and $F_0$ is the
mean response to the optimal grating.  Because results of Fig.~4 are
based only on natural stimuli filters, we have included 5 additional
simple cells for which the natural stimulus filter was reliable and
noise stimulus filter was not.

{\bf Response Reconstruction: Neural Filters and Corresponding
  Nonlinearities.}  In the framework of the LN model, the probability
of response to a particular input {\bf S} is given by an arbitrary
nonlinear function $f$ which only depends on the product of the input
signal {\bf S} and the neural filter {\bf L}:
\begin{equation}
f=f({\bf L*S}).  
\end{equation}
More generally, reconstruction might require description in terms of a
nonlinear function of the outputs of several filters, or curved
subspaces instead of a strictly linear projection between signals and
filters. However, in this paper we focus on the analysis of properties
of the dominant filter {\bf L} of the LN model obtained with noise or
natural inputs. We note that the assumption of a single linear filter
is more general than the assumption that the cell is linear overall,
because the input/output function can be strongly nonlinear and is
usually well described by a threshold or threshold-linear function.

In the case of white noise inputs, the linear filter can be found
using the reverse correlation method, also known as the
spike-triggered average (STA): 
\begin{equation}
\hat e_{\rm STA}=\langle {\bf S} P\left({\rm spike}|{\bf S}\right)\rangle-P\left({\rm spike}\right)\langle{\bf S}\rangle, 
\end{equation} where the expectations are taken
over the stimulus ensemble probability distribution $P({\bf S})$. In
other words, the STA vector is computed by taking the average stimulus
weighted by the number of spikes it elicits and subtracting the
average stimulus multiplied by the overall number of spikes. The
magnitude of the filter is irrelevant, because its change can be
accommodated by an appropriate rescaling of the input-output function
(2), which converts stimulus components along the relevant filter into
spike probability. Therefore, we normalize all of the derived filters
to unit length or measure them with respect to the noise level.

If inputs are taken from a Gaussian distribution with correlations
(colored noise), then the linear filter can be estimated by computing
the STA according to Eq. (3) with a subsequent correction for input
correlations. The decorrelated STA (dSTA) is obtained by multiplying
the STA with the inverse of the stimulus covariance matrix $C_{ij}$:
\begin{equation}
{\hat e}_{\rm dSTA}=C^{-1}{\hat e}_{\rm STA}
\end{equation}
In the case of correlated Gaussian inputs, the dSTA filter Eq. (4)
represents the solution of both the purely linear model and the LN
model. This is no longer true for natural inputs, which are not
Gaussian\cite{Sharpee04}. Therefore we calculate and treat the dSTA
for the natural ensemble as the prediction of the purely linear model.
It is known that higher signal-to-noise ratios and smoother filters
can be achieved by various forms of regularization of the
decorrelation process, including low-pass filtering the STA or
imposing a high-frequency cutoff on the covariance
matrix\cite{Theunissen01,Smyth03,Felsen05}.  The increase in
predictive power upon such regularization happens for three reasons.
First, due to finite data or simply the nature of the stimulus
ensemble, the covariance matrix might be singular or nearly so, so
that its inversion would result in uncontrollably large eigenvalues
for high frequencies where power in the stimulus ensemble is small.
We have found that this is not the case for our covariance matrix:
calculation of the dSTA according to Eq. (4) without any
regularization, in numerical simulations for model linear cells, led
to excellent agreement between the dSTA and the filter of the model
cell with correlation coefficients $>0.99$\cite{Sharpee04} (and
unpublished data).  Second, due to finite amounts of data, there is
noise in the estimation of the STA. If this noise has a relatively
flat spectrum, then at high frequencies where signal in the true STA
is low, decorrelation may preferentially amplify noise rather than
signal.  Again, our results with the linear model with a finite number
of spikes (e.g. 1000 spikes) suggest that this is not a problem,
although we cannot be certain that the noise problem is not worse for
real nonlinear neurons. Third, because the dSTA is a biased estimate
of the filter of an LN neuron probed with natural scenes, the estimate
might be improved by deviating from the linear model.  This can be
done by adding a parameter (a low-pass cutoff) and tuning this
parameter on a cell-by-cell basis to maximize predictive power of the
resulting filter\cite{Theunissen01,Smyth03}.  However, it is not clear
to what degree a change in just one parameter could account for all
deviations between filters of the fully linear model and those of the
LN framework. For all of these reasons, we refrained from
regularization in our calculations of the dSTA except in the
illustrations of example cells in Fig.~1; we otherwise treated the
dSTA calculated by Eq. (4) as the prediction of the fully linear
model. It should also be noted that the inclusion of an ad-hoc
low-pass filter parameter would make it impossible to reliably
estimate the higher-frequency parts of the filter; this, along with
the bias of the unregularized dSTA, is why the MID method was
necessary for us to assay changes in the spatial frequency tuning
across ensembles.  In Fig.~1, for comparison purposes, we illustrate
both regularized and unregularized forms of the dSTA.  Regularization
was based on selecting a cutoff on the eigenvalues of the covariance
matrix $C$ below which none of the eigenvalues with the corresponding
eigenvectors contributed to the inverse $C^{-1}$ in Eq. (4), making it
a pseudo-inverse\cite{Theunissen01,Smyth03}. For each possible value
of the cutoff parameter, the dSTA vector was calculated according to
Eq. (4) based on a trial set using 7/8 of the data. The optimal
cutoff value was selected as that for which the corresponding dSTA
provided maximal information on the remaining 1/8 of the data
designated as a test set.

In addition to the above methods, we also derived neural filters using
the method of most informative dimensions\cite{Sharpee04}, see next
section. For all of the above methods, jackknife analysis of neural
filters was performed: 8 filters were computed, each with 1/8 of the
data left out. When information was computed for a filter on its own
ensemble, it was calculated only on this $1/8$ of the data that was not
used for computing the filter, except in Fig.~4 where a single
filter was calculated from $1/2$ of the data and information was
calculated on segments of the other half. In all other cases,
information values reported are an average over the 8 values found
with the 8 jackknife estimates. To establish statistical significance
of the difference between filters derived with any two different
methods and/or two stimulus ensembles, all 16 of the corresponding
jackknife estimates (8 for each combination of method and ensemble)
were projected on the direction of the difference between the mean
filters describing the two groups, and an unpaired Students $t$-test was
used on these projections. To calculate the signal-to-noise level of
receptive fields shown in Fig.~1, we compute the average standard
deviation across all components of the receptive field across all the
jackknife estimates (normalized to unit length) and display receptive
field values relative to that noise level.

\begin{figure*}
\includegraphics[width=5.5in]{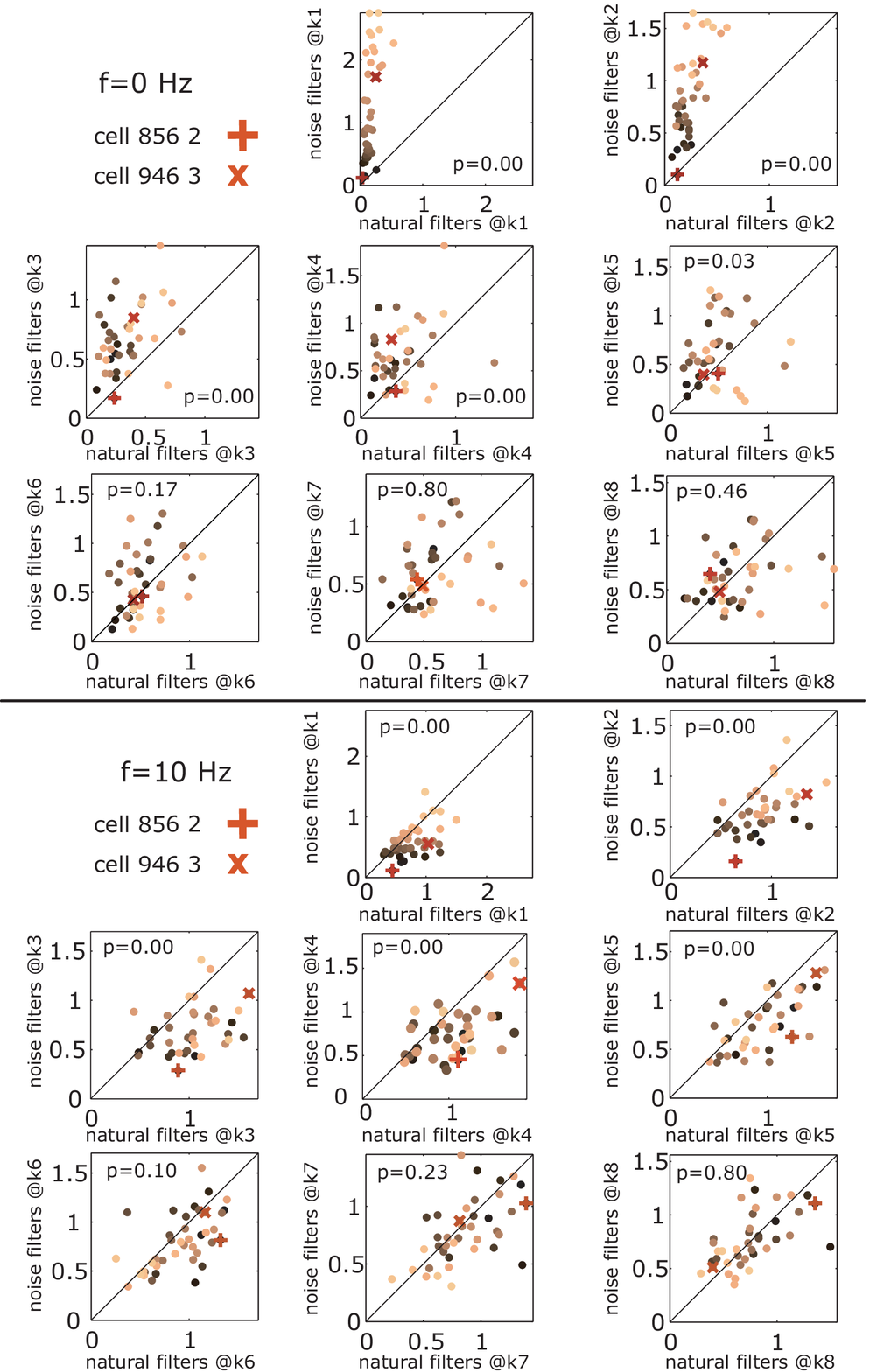}
\parbox{7in}{\begin{flushleft}
SUPPLEMENTARY FIG. 2: Spatial frequency sensitivity on a
cell-by-cell basis for the first 9 spatial frequencies from Fig.~2
(here called k1 to k8 from lowest to highest) for temporal frequencies
of 0 and 10 Hz respectively. P-values on top of each graph show
significance in sensitivity differences of filters derived from noise
vs. natural stimulation. Color for each cell codes sensitivity to
noise filter at lowest frequency (k1) and is retained in the plots of
higher frequencies. The two example cells of Supplementary Fig.~1
are marked as a '+' and an 'X' respectively.  Note that the cell
marked by a '+' is atypical in its behavior at 0Hz.
\end{flushleft}
}
\end{figure*}

Once the filter {\bf L} has been obtained as either the STA (3), dSTA
(4), or the MID\cite{Sharpee04}, we can calculate the nonlinear
input-output function (2) directly from the data. According to its
definition it is given by the normalized spike probability given the
stimulus {\bf S}:
\[
f({\bf S}*{\bf L})=\frac{P({\rm spike}|{\bf S})}{P({\rm spike})}.
\]
When working in the framework of the linear-nonlinear model we assume
that the spike probability only depends on stimulus components along
the filter {\bf L} of interest: $ P({\rm spike}|{\bf S})=P({\rm
  spike}|{\bf S*L})$. Therefore the nonlinear input/output function
can also be written as:
\[
f({\bf S}*{\bf L})=\frac{P({\rm spike}|{\bf S}*{\bf L})}{P({\rm spike})}.
\]
The last expression can be transformed using Bayes' rule: 
\begin{equation} 
f({\bf S}*{\bf L})=\frac{P({\bf S}*{\bf L}|{\rm spike})}{P({\bf S}*{\bf L})}.
\end{equation} 
That is, the nonlinear input/output function $f$ is evaluated as a
ratio of probability distributions of stimulus components along the
filter {\bf L}, $P({\bf S*L})$, and of the probability distribution of
stimulus components $P({\bf S*L}|{\rm spike})$ conditional on a spike.
Both of the probability distributions are readily available from the
experimental data.  

{\bf Reconstruction of Receptive Fields as Most Informative
  Dimensions.} The justification for the method of most informative
dimensions as a way to calculate neural receptive fields is described
elsewhere\cite{Sharpee04}, where performance of the method is
illustrated on model visual and auditory neurons. For the convenience
of the reader we describe here the methodology of maximizing
information to find the receptive fields.  It was shown that the
information between the output of a particular vector {\bf L} in the
input space and the neuron's response, regarded as a spike or no spike
in each time bin, can be computed, to lowest order in the probability
$P({\rm spike})$ of a spike in the time bin, as the Kullback-Leibler
distance between the probability distributions $P(x)$ and $ P(x|{\rm
  spike})$: 
\begin{equation}
I({\bf L})=\int dx P_{\bf L}(x|{\rm spike}){\rm log}_2\left[ \frac{P_{\bf L}(x|{\rm spike})}{P_{\bf L}(x)}\right],
\end{equation} 
where $P_{\bf L}(x)$ is the probability distribution of stimulus
projections $x$ onto the vector {\bf L} in the input ensemble, and
$P_{\bf L}(x|{\rm spike})$ is the probability distribution of stimulus
projections $x$ onto the vector {\bf L} among inputs that led to a
spike. We compute these two probability distributions as histograms in
21 bins covering the range of projection values (the same number of
bins was used in finding MIDs from neural responses to noise and
natural ensemble). For each trial vector, we also compute the gradient
of information as: 
\begin{equation}
\nabla_{\bf L}I=\int dx P_{\bf L}(x) \left [ \langle {\bf s} |x, {\rm
    spike}\rangle -\langle {\bf S}|x\rangle\right ] \frac{d}{dx}\left
  [ \frac{P_{\bf L}(x|{\rm spike})}{P_{\bf L}(x)}\right],
\end{equation}
where $\langle{\bf S}|x\rangle$ is the average of the stimuli having
projection value of $x$ onto the vector {\bf L} (using the same
binning of $x$ as for the probability distributions $P_{\bf L}(x)$ and
$P_{\bf L}(x|{\rm spike})$).  Similarly, $\langle {\bf S}|x,{\rm
  spike}\rangle$ is the average of the stimuli that led to a spike
that had projection value of $x$ onto the vector {\bf L}.  We evaluate
the derivative at a particular value of x using Savitsky-Golay
coefficients (W.H.  Press et al., {\it Numerical Recipes}, Cambridge
University Press 1998) based on two adjacent bins on either side of
the bin with the value $x$; if projections values from any one of
these bins were not encountered in the stimulus ensemble, the
corresponding average did not contribute to the derivative. We find
that the use of Savitsky-Golay smoothing coefficients is not required,
but helps improve convergence of the algorithm [note that in the
search algorithm, described below, the trial vectors are accepted
based on information values, which are evaluated without smoothing].
This analysis requires that stimuli and spike trains are binned at the
same time resolution (33 ms for natural stimuli and 16 ms for noise
stimuli). Therefore occasional stimuli correspond to multiple spikes
in a bin. If that happened, projections values of such stimuli were
counted as many times as there were spikes for all the probability
distributions and averages in Eqs. (6) and (7).

\begin{figure*}
\includegraphics[width=5in]{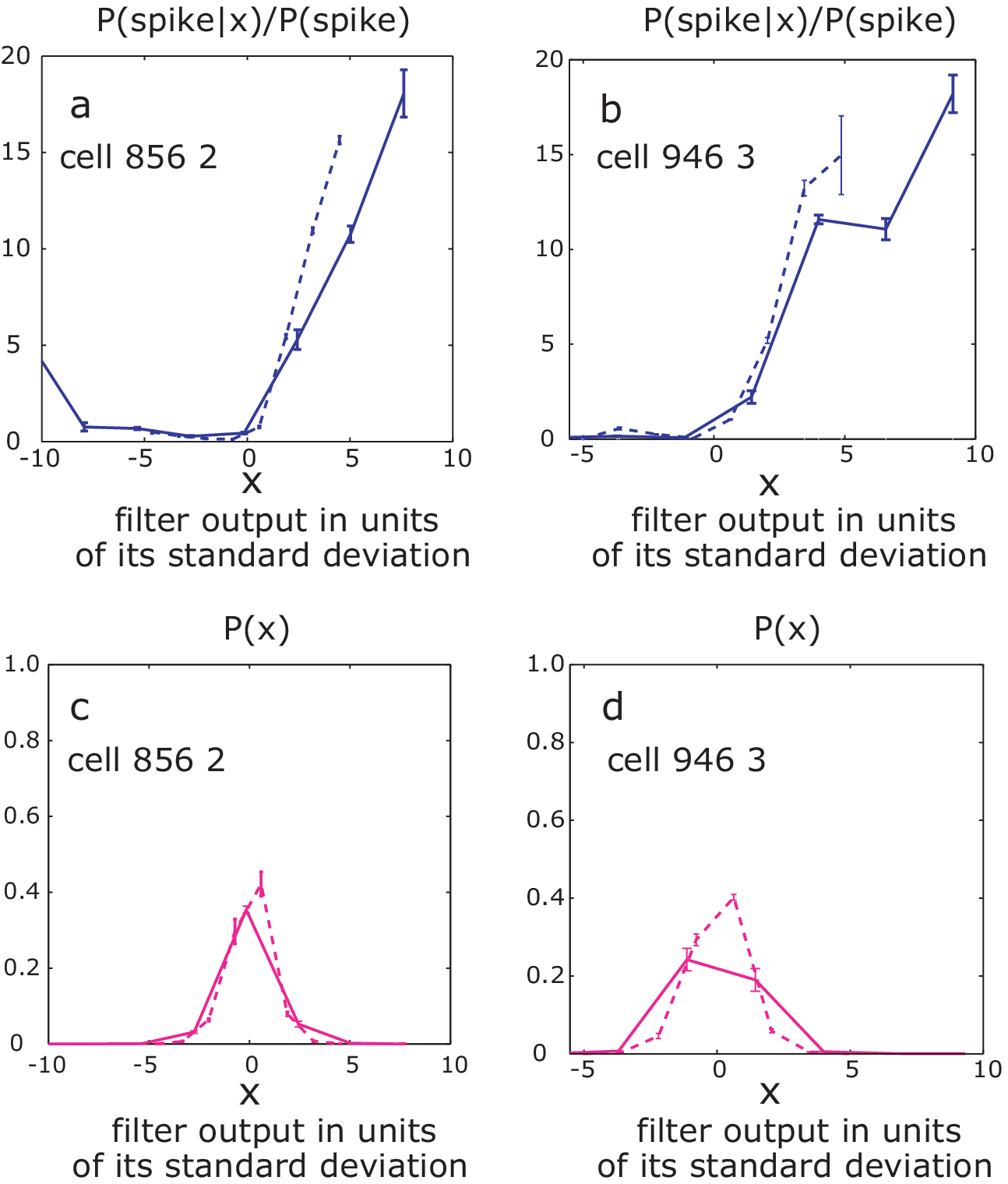}
\parbox{7in}{\begin{flushleft}
SUPPLEMENTARY FIG. 3: Panels (a,b) show that the nonlinear input/output function
  $f(x)=P({\rm spike}|x)/P({\rm spike})$ associated with the MID
  filters for two exemplary cells of Fig.~1 overlap under natural
  (solid) and noise (dashed) stimulation when stimulus projection $x$
  along the corresponding receptive fields is measured in units of its
  standard deviation (x-axis). For comparison, in Fig.~1 we plot the
  input/output function $f(x)$ scaled by the firing rate, $P({\rm
    spike}|x)$ - the probability of a spike in 33ms window given a
  stimulus projection value x along the receptive field. Therefore the
  difference in scale for the nonlinearities observed between natural
  and noise conditions in Fig.~1, as for example cell 856 2,
  reflects only a change in the mean firing rate under the two
  conditions. Panels (c,d) show the probability distributions of
  projections x for natural (solid) and noise (dashed) stimulation.
\end{flushleft}
}
\end{figure*}

The search for the most informative dimension (MID) is initialized by
setting the starting vector equal to the STA. To generate a new trial
vector, we perform a line maximization (W.H. Press et al., {\em
  Numerical Recipes}, Cambridge University Press 1998) along the line
defined by the gradient (7), and choose, on average, the one with the
largest information. Because information (6) as a function of
components of the vector {\bf L} has local maxima, smaller information
values are accepted with Boltzmann probability, $\exp(-\Delta I/T)$,
where $\Delta I$ is the decrease in information between the new and
old trial vector measured in units of the information $I_{\rm spike}$
carried by the arrival of a single spike, and the parameter $T$ is
called the effective temperature of the simulated annealing cooling
scheme.  Information values in these units are typically less than one
(unless there is overfitting). Therefore, we start the simulated
annealing scheme with $T=1$, and decrease it by a factor of 0.95 after
each line maximization.  If the search appears to have converged with
a fraction precision of $5 \times 10^{-5}$ and the effective temperature
$T \leq 10^{-5}$, then the effective temperature is increased by a factor
of 5, but not to exceed the starting temperature value. This results
in repeated "cooling" and "remelting", and is equivalent to restarting
the algorithm multiple times. We limit the total number of line
maximizations to 3000. The best vector found in terms of information
during the overall maximization procedure is taken as the most
informative dimension {\bf L}.  Cross-validation is performed by
leaving out 1/8 of data and treating that 1/8 as a test set. We
compute information on the test set after every 100 line
maximizations, and if the information value has dropped on the test
set by 25\% of its maximum value, the optimization procedure is
stopped and the current filter taken as the MID. Such early stopping
seldom occurs when we compute receptive fields from responses to
natural scenes, but is common when receptive fields are computed from
noise ensembles. This is due to the fact that the starting point, the
STA, is very close to the optimal value when neural responses to the
noise ensemble are analyzed.

\begin{figure*}
\includegraphics[width=5in]{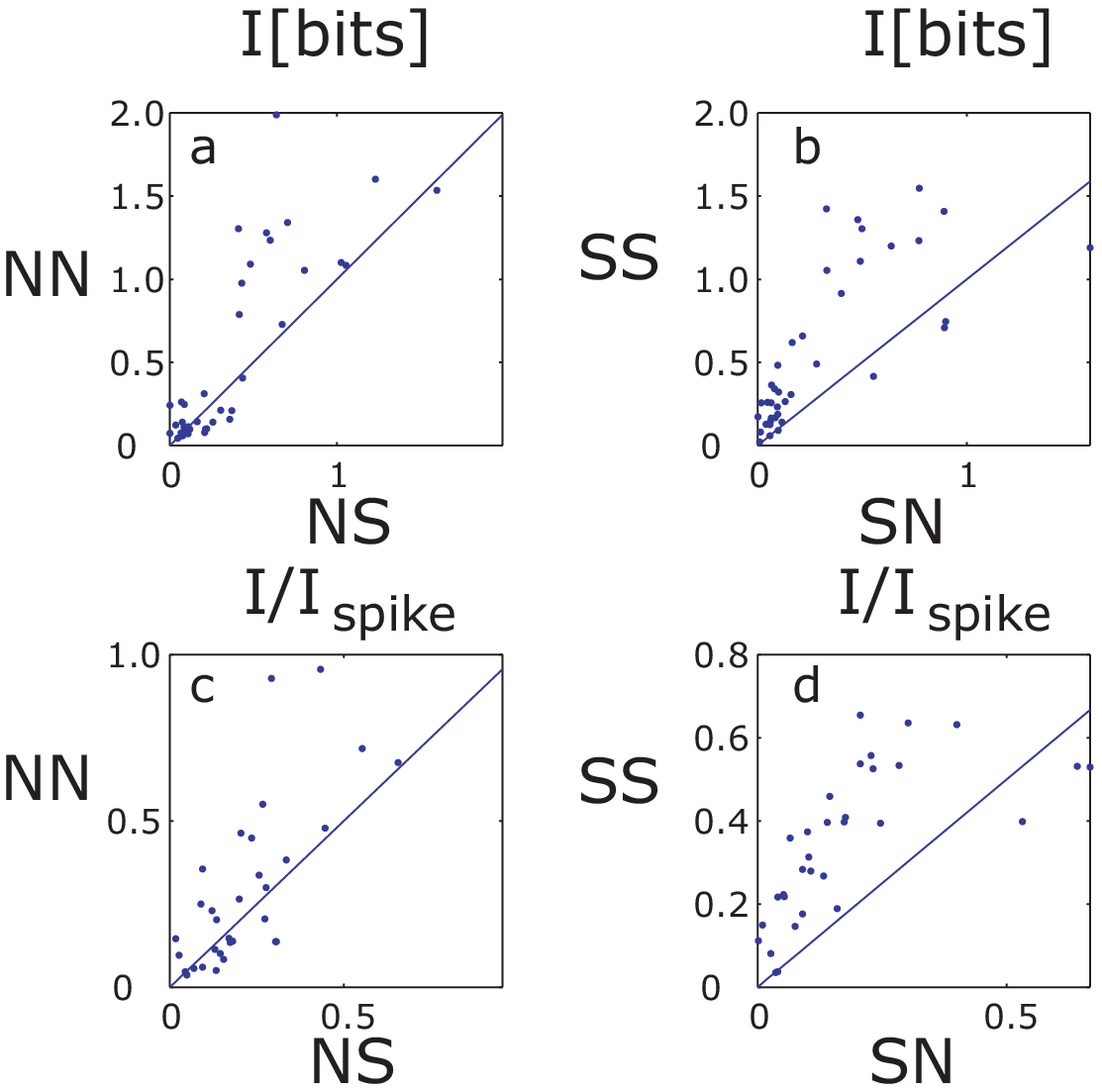}
\parbox{7.0in}{
\begin{flushleft}
  SUPPLEMENTARY FIG. 4: The increase in information on a cell-by-cell
  basis when the noise filter is applied to the noise vs. natural
  ensemble (a) or when the natural filter is applied to the natural
  vs. noise ensemble (b). Panels (c,d) show this effect in units of
  $I_{\rm spike}$.  Notations are as in Fig.~3.
\end{flushleft} }
\end{figure*}

Because the MID method is based on a search in a high-dimensional
space for an information maximum, there is of course a concern that
our search might become stuck in a local maximum.  We believe this is
not a concern for the following reasons.  First, as just noted, our
search procedure is equivalent to restarting the search algorithm
multiple times from multiple starting points, only the first of which
is the STA, and we take the maximum of information over the entire
search.  Second, in studies of model cells\cite{Sharpee04}(and
unpublished data), we have found that the error (measured as 1 minus
the projection between the true model filter and the MID found by the
search) decreases as $1/N$ where $N$ is the number of spikes used to
estimate the filter.  This is the dependence predicted
theoretically\cite{Sharpee04}, and would not be expected to hold if
the true maximum were not being found.  Third, we have previously
verified on model cells that beginning with a random starting point
rather than the STA does not produce better solutions.  The STA
represents a natural choice of a starting point in that it is clearly
a stimulus direction that carries nonzero information about the
neuron's response.

{\bf The MID method produces an unbiased estimate.}  In this section
we provide a detailed derivation for the fact, first published in
Ref.\cite{Sharpee04}, that the MID method produces unbiased estimates
of neural filters within a single-filter LN model. We will first
consider the case of infinite data, and then go through details of the
argument with finite data.

While the MID filter can be calculated with respect to any particular
pattern of spikes\cite{Sharpee04}, in this paper we have concentrated
on finding filters associated with single spikes. Therefore we will do
so in this section as well. Information carried by individual spikes
about the incoming stimuli is given by\cite{B00a}: 
\begin{equation}
I_{\rm spike}=\int d^D{\bf S} P({\bf S}) \frac{P({\rm spike}|{\bf
    S})}{P({\rm spike})}{\rm log}_2\frac{P({\rm spike}|{\bf
    S})}{P({\rm spike})}.
\end{equation}
Because this is the information between single spikes and full,
unfiltered, stimuli, information between spikes and stimuli filtered
along any dimension may not exceed (8). To verify that the only filter
that leads to an equal amount of information between spikes and
stimuli filtered with it is the neural receptive field {\bf L}, we
invoke the main assumption of the single-filter LN model: $P({\rm
  spike}|{\bf S})=P({\rm spike}|{\bf S*L})$, so that:
 \[
I_{\rm spike}=\int d^D{\bf S} P({\bf S}) \frac{P({\rm spike}|{\bf
    S*L})}{P({\rm spike})}{\rm log}_2\frac{P({\rm spike}|{\bf
    S*L})}{P({\rm spike})}.
\]
The integration $d^D{\bf S}$ with along all stimulus dimensions can be
carried out separately along the relevant stimulus dimension, {\bf
  S*L}, and along the rest of stimulus dimensions, which we denote as
${\bf S}_\perp$:

\begin{eqnarray*}
  I_{\rm spike}&=&\int d({\bf S*L}) \frac{P({\rm spike}|{\bf
      S*L})}{P({\rm spike})}{\rm log}_2\frac{P({\rm spike}|{\bf
      S*L})}{P({\rm spike})}\nonumber \\ && \times \int d^{D-1}{\bf S}_\perp P({\bf S}_\perp,
  {\bf S*L}) 
\end{eqnarray*}
Integration with respect to all of the irrelevant stimulus dimensions
${\bf S}_\perp$ results in:
  \begin{eqnarray*}
I_{\rm spike}&=&\int d({\bf S*L}) \frac{P({\rm spike}|{\bf
    S*L})}{P({\rm spike})}{\rm log}_2\frac{P({\rm spike}|{\bf
    S*L})}{P({\rm spike})} \nonumber \\ && \times P({\bf S*L}), 
\end{eqnarray*}
which is precisely the information along the filter {\bf L}, cf. Eq.
(6). We have thus shown that information along the filter that
represents the neural receptive field achieves the maximal information
possible, $I_{\rm spike}$ and describes the encoding ${\bf S}\to {\bf
  S*L}\to {\rm spikes}$. Filtering along any other dimension {\bf V}
will correspond to encoding ${\bf S}\to {\bf S*V} \to {\bf S*L}\to
{\rm spikes}$ or ${\bf S}\to {\bf S*L} \to {\bf S*V }\to {\rm spikes}$
and, by the data processing inequality (Cover and Thomas, John Wiley
Inc. 1991), leads to a lower information processing value. The data
processing inequality applies to stochastic inputs but presumes that
we know exact probabilities such as $P({\bf S*L}| {\rm spike})$ and
$P({\bf S*V}| {\rm spike})$.  This shows that the MID method is
unbiased in the limit of infinite data and stochastic neurons.

\begin{figure*}
\includegraphics[width=6.5in]{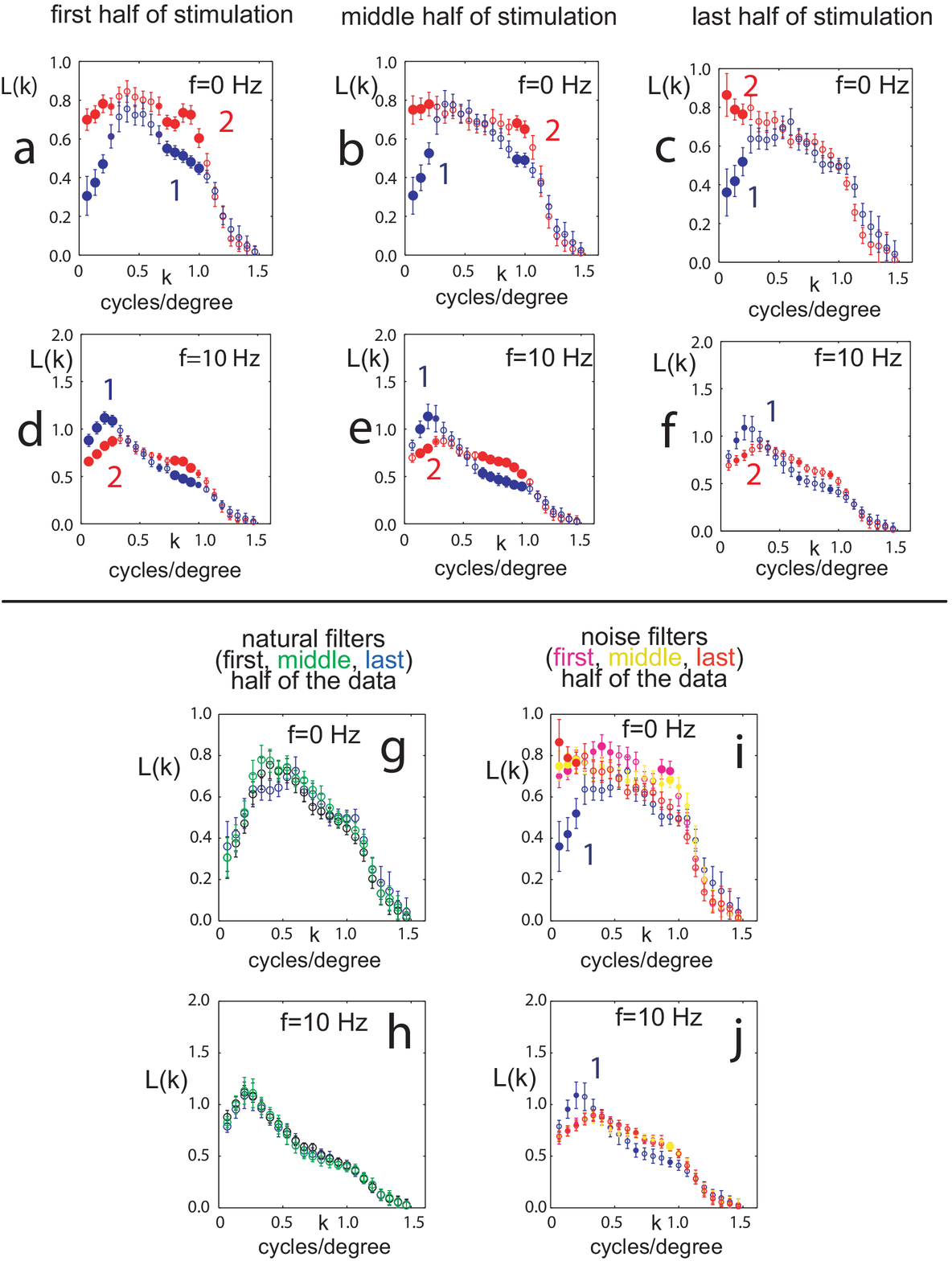}
\parbox{7.in}{\begin{flushleft}
SUPPLEMENTARY FIG. 5: {\bf Coarse evolution of adaptive neural filters.} (a,d)
  Comparison of neural filters derived from the first half (a,d),
  middle half (b,e) or last half (c,f) of stimulation with noise and
  natural inputs. Notations are as in Fig.~2(a,d).  In panels (g,h)
  we plot only natural filters to show that they overlap. In panels
  (i,j) we compare three of the noise filters derived from the first
  half of the data (magenta), middle half of the data (yellow), and
  last of the data (red) to the natural filters of the last half of
  the data. With time, noise neural filters diverge from natural
  filters.
\end{flushleft}}
\end{figure*}

With finite data, we have only a limited number of samples to measure
the probability distributions $P({\bf S*L}| {\rm spike})$ and $P({\bf
  S}|{\rm spike})$. With $N$ spikes, our empirical estimates of these
probability distributions $P_N({\bf S*L}|{\rm spike})$ and $P_N({\bf
  S}| {\rm spike})$ will differ from experiment to experiment in such
a way that the average across trials produces the true distribution
and the variance across trials acquires a term of the order of $1/N$:
\begin{eqnarray}
  \langle P_N({\bf S}|{\rm spike})\rangle&=&P({\bf S}|{\rm spike}) \\
   \langle P_N({\bf S}|{\rm spike})^2\rangle&=&P({\bf S}|{\rm
    spike})^2+\frac{1}{N}P({\bf S}|{\rm spike}) \nonumber \\
&&\times \left(1-P({\bf S}|{\rm spike})\right)
\end{eqnarray}
where we have used the properties of the binomial distribution; each
particular stimulus {\bf S} can occur with a spike anywhere between 0
and $N$ times, if $N$ is the total number of spikes.  Similar
relations can be used with other probability distributions involved.

The deviation between the true filter and the MID filter obtained with
a particular data set, $\delta {\bf V}$, is proportional to the
gradient of information (evaluated with finite data) at the position
of the true filter: $\delta {\bf V}\sim \nabla I({\bf L})$.  Here we
show that, as was stated in Ref.\cite{Sharpee04}, the gradient of
information is zero, after averaging across trials, for the true
filter. To verify this we represent information $I_N({\bf L})=I({\bf
  L})+\delta I_N({\bf L})$, as the information obtained with infinite
data and the deviation from it due to finite sampling.  The gradient
of the information is zero at the true filter {\bf L}.  The deviation
\begin{eqnarray}
  \delta I_N({\bf L}) &=& \int dx \delta P_N(x|{\rm spike}) {\rm
    log}_2\left [ \frac{P(x|{\rm spike})}{P(x)}\right ]
\nonumber \\
&&+\int dx \delta  P_N(x|{\rm spike}),
\end{eqnarray}
where $x={\bf S*L}$, $\delta P_N(x|{\rm spike})=P_N(x|{\rm
  spike})-P(x|{\rm spike})$ is the difference between the empirical
and true distributions, and there is no need to consider noise in the
stimulus distribution $P(x)$ because it might be taken as the one
actually used in the experiment. Next we take into account that the
empirical distribution obeys a normalization constraint, such that
$\int dx P_N(x|{\rm spike})=1$, and therefore $\int dx \delta
P_N(x|{\rm spike})=0$, so that: 
\begin{equation}
\delta I_N({\bf L})=\int dx \delta P_N(x|{\rm spike}){\log}_2\left[
\frac{P(x|{\rm spike})}{P(x)}
\right],
\end{equation}
But the average of the empirical distributions is the true
distribution (9), so $\delta I_N({\bf L})=0$ in the first-order
approximation in the deviations between empirical and true
distributions. The second-order approximation results, using the
property Eq. (10), in a uniform correction: $\delta I_N({\bf L})\sim
\frac{N_{\rm bins}-1}{N_{\rm spike}}$, where $N_{\rm spike}$ is the
number of spikes and $N_{\rm bins}$ is the number of bins used in
estimating the probability distribution $P(x|{\rm spike})$.  Because
this correction is independent of the direction in the stimulus space,
it provides a zero contribution to the gradient at the position of the
true filter. The second-order terms determine the variance of the MID
filters on a trial-by-trial basis, because while the deviations
themselves $\delta {\bf V}\sim \nabla I({\bf L})$ are proportional to
the gradient of information, their variance $ \langle \delta {\bf V}_i
\delta {\bf V}_j\rangle \sim \langle \nabla_i I({\bf L})\nabla_j
I({\bf L})\rangle $ is proportional to pairwise gradient correlations.
Using Eqs.  (11) and (10), one can show that the leading term
determining this variance behaves as $\dim 1/N_{\rm spike}$. The exact
coefficient can be found in Ref.  \cite{Sharpee04}. This means that
while different MID filters obtained based on different empirical
distributions deviate from each other and from the true filter, these
deviations have zero mean and finite variance that decreases as $\sim
1/N_{\rm spike}$ with increasing number of spikes.  While there may be
terms $\sim N_{\rm spike}^{-2}$ describing a shift in the mean, these
will be masked by a much larger effect of variance between estimates
decreasing as $N_{\rm spike}^{-1}$. This is what we mean by saying
that the MID method is unbiased. Note that the gradient of information
evaluated at the filters of the linear model (STA or decorrelated STA)
will be non-zero, with terms of order $O(1)$, which do not depend on
the number of spikes and remain finite even in the limit of infinite
data.

\begin{figure*}
\includegraphics[width=5in]{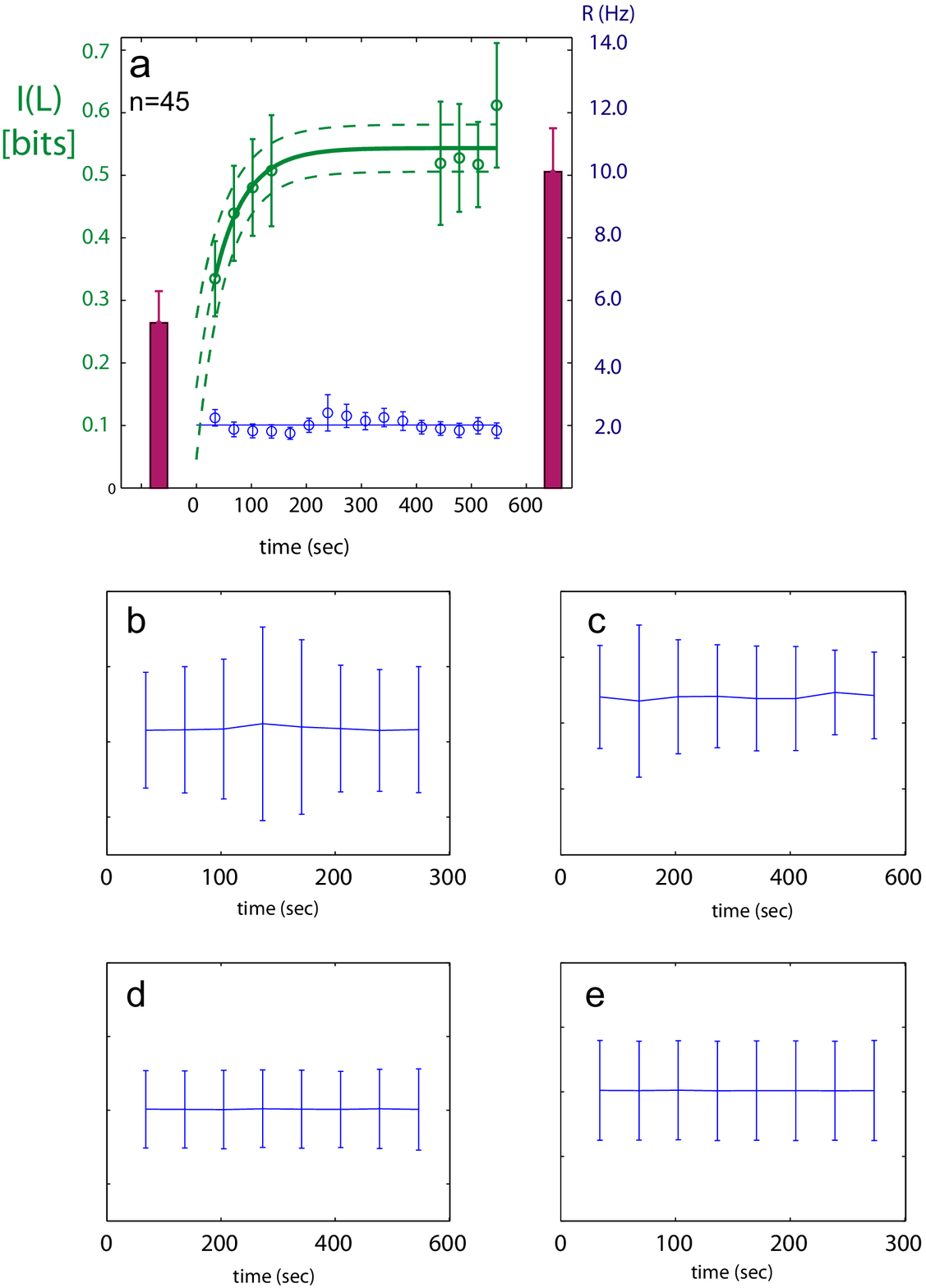}
\parbox{7in}{
\begin{flushleft}
  SUPPLEMENTARY FIG. 6: (a) The neural filter derived from the middle
  half of natural stimulation is applied to the first and last quarter
  of the natural input ensemble. Notations are as in Fig.~4. The solid
  line is an exponential fit, dashed lines show one standard deviation
  based on the Jacobian of the fit, $p=0.007$. The remaining panels
  show that the relevant statistical properties of the input ensemble
  are stable and cannot account for the time dependence seen in
  Fig.~4. Here we show the mean and standard deviations (in arbitrary
  units) for natural and noise input stimuli filtered differently: (b)
  natural stimuli (first half of the data) filtered with natural
  neural filters computed from second half of the data; (c) natural
  stimuli (all duration) filtered with noise neural filters; (d) noise
  stimuli (all duration) filtered with natural neural filters; (e)
  noise stimuli (first half of the data) filtered with noise neural
  filters obtained from the second half.
  \end{flushleft}}
\end{figure*}

\begin{figure*}
  \includegraphics[width=5in]{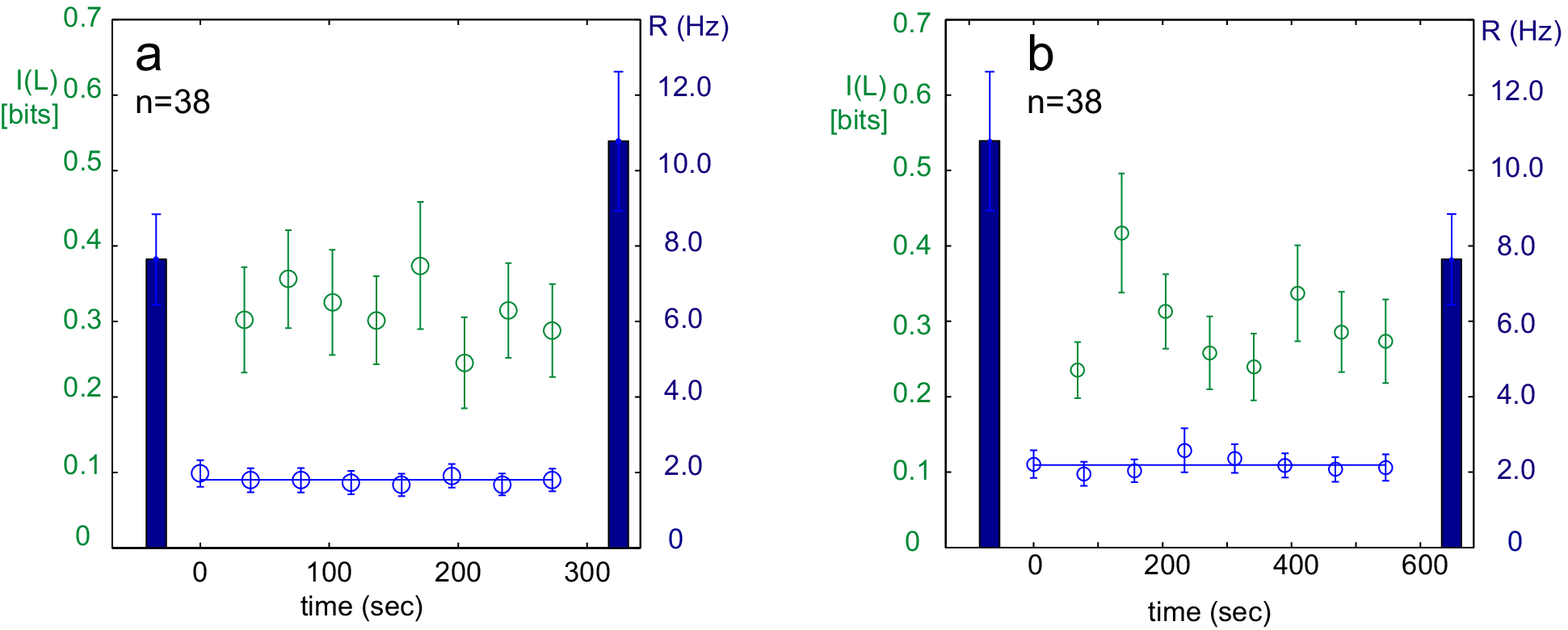} 
\parbox{7.0in}{
    \begin{flushleft}SUPPLEMENTARY FIG. 7: Information carried by the
      noise filter about the neuron's response, as a function of time
      after exposure to the noise ensemble (a) or natural stimulus
      ensemble (b). Information values were evaluated along the noise
      filter derived from the second half (a) and from full recording
      (b) of noise stimulation. No significant time dependence could
      be established. Notations are as in Fig.~4.  Left and right
      blue bars show average information carried by noise filter about
      responses to noise ensemble (taller bar) or natural ensemble
      (shorter bar). Note that the average information values computed
      for the short time segments for the noise filter applied to the
      noise ensemble (a) are all smaller than the average information
      computed over the whole noise ensemble (right bar in a).  This
      suggests that these short-time estimates are too noisy to be
      reliable in the case of the noise filter, which may provide
      another reason that we could observe no trend for the noise
      filter.  A similar problem can be seen in (b). Note that a
      similar problem did not arise for the natural filter (main text,
      figure 4): short-time estimates were equal in size to the
      estimate over the whole ensemble after adaptation.  We used the
      filter from the full recording in (b) (unlike in main text,
      figure 4, where the same filter was used in (a) and (b) for
      consistency) because the short-time estimates for the filter
      from the second half of the recording showed an even stronger
      tendency to have low information values; using the full
      recording helps fight noise and so improves the situation, but
      not sufficiently.
\end{flushleft}
}
\end{figure*}

However there are ways in which the stimulus ensemble can influence
the single MID even in a neuron that does not adapt, if the relevant
subspace (RS) has two or more dimensions. In this case, as shown in
Ref.~\cite{Sharpee04}, Appendix B, the single MID for that ensemble
may include a component outside of the RS if the ensemble is such that
the average stimulus given the projections along the relevant
dimensions is not a linear function of each projection (as can occur
for non-Gaussian ensembles). Any such effects, however, would be
instantaneous and would not yield a time-dependence to the calculation
of information as in Fig. 4.

{\bf Details of stimulus presentation and filter analysis.} The visual
input signals were presented as two-dimensional spatiotemporal
patterns of light intensities on a video monitor with a refresh rate
of 120 Hz.  The frame update rate was 60 Hz in the case of the white
noise stimulus ensemble and 30 Hz in the case of the natural stimulus
ensemble (our commercial cameras did not provide higher temporal
resolution than that of television, which is 30Hz). No corrections
were made for the camera nonlinear amplitude to intensity
transformation function.

The optimal orientation was determined from responses to a set of
evenly spaced orientations at 10° intervals, with a spatial frequency
of 0.5 cycles/degree and a temporal frequency of 2 Hz. The optimal
spatial frequency was derived from responses to a set of moving
gratings of optimal orientation and variable spatial frequencies
(approximately logarithmically spaced between 0.1 and 4
cycles/degree).

Spatial frequency profiles were obtained by taking the Fourier
transform in time and, with zero-padding to 32x32, in space. Linear
interpolation between pixels of the 2D transform was used to derive
one-dimensional profiles along the preferred orientation of each cell.
Before averaging across cells, the spatial frequency profiles of
individual cells were normalized to unit length across all spatial and
temporal frequencies. Identical procedures were used for receptive
fields and stimuli comprising the input ensembles (averaging over all
three frame subsequences, e.g. 1-2-3, 2-3-4, etc.).

In Fig.~3, the information $I$ was calculated from jackknife
estimates of the filters. For each cell, for either the natural or
noise ensemble, eight jackknife estimates were derived, each from 7/8
of the data with the remaining 1/8 of the data serving as a test set
on which the information was calculated. The mean of these 8 estimates
was assigned as information $I$ that cell and ensemble.  $I_{\rm
  spike}$ is calculated from responses to 50-150 repetitions of an
11s-long segment of the natural or noise ensemble. Finite-size
corrections\cite{B00a} were applied to both $I$ and $I_{\rm spike}$.
As a control for the information calculation, we calculated natural
MID filters for a series of model simple cells with a static filter
where the number of spikes emitted over the course of the test set
varied from 80-13,000. The calculated information, of course,
decreased substantially at low numbers of spikes, but it did so
similarly whether the filter was applied to the natural or the noise
ensemble. There was no significant difference between the information
about the natural ensemble and about the noise ensemble for any choice
of nonlinearity, that is for any signal-to-noise ratio.

\end{document}